\documentclass[useAMS,usenatbib]{mn2e}

\usepackage{graphicx}	% Including figure files
\usepackage{amsmath}	% Advanced maths commands
\usepackage{amssymb}	% Extra maths symbols
\usepackage{multicol}        % Multi-column entries in tables
\usepackage{bm}		% Bold maths symbols, including upright Greek
\usepackage{pdflscape}	% Landscape pages
\usepackage{multirow,multicol} 

\def\N1316{NGC\,1316}
\def\N1404{NGC\,1404}

\def\4U{4U~1735$-$444}

\def\arcsec{\ifmmode '' \else $''$\fi}

\def\arcsecpoint{\ifmmode ''\!. \else $''\!.$\fi}

\def\kms{\ifmmode {\rm km\ s}^{-1} \else km s$^{-1}$\fi}
\def\Msun{\ifmmode {\rm M}_{\odot} \else M$_{\odot}$\fi}
\def\Lsun{\ifmmode {\rm L}_{\odot} \else L$_{\odot}$\fi}
\def\Zsun{\ifmmode {\rm Z}_{\odot} \else Z$_{\odot}$\fi}

\def\ergscm2{ergs\,s$^{-1}$\,cm$^{-2}$}
\def\icm3{{\rm cm}^{-3}}
\def\icm2{{\rm cm}^{-2}}
\def\qo{\ifmmode q_{\rm o} \else $q_{\rm o}$\fi}
\def\Ho{\ifmmode H_{\rm o} \else $H_{\rm o}$\fi}
\def\ho{\ifmmode h_{\rm o} \else $h_{\rm o}$\fi}

\def\vFWHM{\ifmmode v_{\mbox{\tiny FWHM}} \else
            $v_{\mbox{\tiny FWHM}}$\fi}
\def\CCF{\ifmmode F_{\it CCF} \else $F_{\it CCF}$\fi}
\def\ACF{\ifmmode F_{\it ACF} \else $F_{\it ACF}$\fi}
\def\Halpha{\ifmmode {\rm H}\alpha \else H$\alpha$\fi}
\def\Hbeta{\ifmmode {\rm H}\beta \else H$\beta$\fi}
\def\Hgamma{\ifmmode {\rm H}\gamma \else H$\gamma$\fi}
\def\Hdelta{\ifmmode {\rm H}\delta \else H$\delta$\fi}
\def\Lya{\ifmmode {\rm Ly}\alpha \else Ly$\alpha$\fi}
\def\Lyb{\ifmmode {\rm Ly}\beta \else Ly$\beta$\fi}
\def\Lyg{\ifmmode {\rm Ly}\beta \else Ly$\gamma$\fi}

\def\ciii{\ifmmode {\rm C}\,{\sc iii} \else C\,{\sc iii}\fi}
\def\civ{\ifmmode {\rm C}\,{\sc iv} \else C\,{\sc iv}\fi}
\def\cv{\ifmmode {\rm C}\,{\sc v} \else C\,{\sc v}\fi}
\def\cvi{\ifmmode {\rm C}\,{\sc vi} \else C\,{\sc vi}\fi}

\def\o5007{[O\,{\sc iii}]\,$\lambda5007$}

\def\fexxii-iii{Fe\,{\sc xxii-xxiii}}

\usepackage{hyperref}

\hypersetup{colorlinks=true,linkcolor=blue,citecolor=blue,filecolor=blue,urlcolor=blue}

\title[Thermal stability of super-Eddington winds]{\centering{Thermal stability of winds driven by radiation pressure \\ in super-Eddington accretion discs}}
\author[C. Pinto et al.]{C. Pinto,$^{1,2,3}$\thanks{E-mail:
ciro.pinto@esa.int} M. Mehdipour,$^{4}$ D. J. Walton,$^{2}$ M. J. Middleton,$^{5}$ T. P. Roberts,$^{6}$ 
\newauthor A. C. Fabian,$^{2}$ M. Guainazzi,$^{1}$ R. Soria,$^{7,8}$ P. Kosec\,$^{2}$ and J.-U. Ness\,$^{9}$  \\
$^{1}$ESTEC/ESA, Keplerlaan 1, 2201AZ Noordwijk, The Netherlands\\
$^{2}$Institute of Astronomy, Madingley Road, CB3 0HA Cambridge, United Kingdom\\
$^{3}$INAF - IASF Palermo, Via U. La Malfa 153, I-90146 Palermo, Italy\\
$^{4}$SRON Netherlands Institute for Space Research, Sorbonnelaan 2, 3584 CA Utrecht, The Netherlands\\
$^{5}$Physics \& Astronomy, University of Southampton, Southampton, Hampshire SO17 1BJ, UK\\
$^{6}$Centre for Extragalactic Astronomy, Durham University, Department of Physics, South Road, Durham DH1 3LE, UK\\
$^{7}$College of Astronomy and Space Sciences, University of the Chinese Academy of Sciences, Beijing 100049, China\\
$^{8}$Sydney Institute for Astronomy, School of Physics A28, The University of Sydney, Sydney, NSW 2006, Australia\\
$^{9}$ESAC/ESA European Space Astronomy Center, P.O. Box 78, 28691 Villanueva de la Canada, Madrid, Spain}

\begin{document}

\date{\today}

\pagerange{\pageref{firstpage}--\pageref{lastpage}} \pubyear{2018}

\maketitle

\label{firstpage}

\begin{abstract}
Ultraluminous X-ray sources (ULXs) are mainly powered by accretion in neutron stars or stellar-mass black holes. Accreting at rates exceeding the Eddington limit by factors of a few up to hundreds, radiation pressure is expected to inflate the accretion disc, and drive fast winds that have in fact been observed at significant fractions of the speed of light.
Given the super-Eddington luminosity, the accretion disc will be thicker than in sub-Eddington accretors such as 
common active galactic nuclei and X-ray binaries, leading to a different spectral energy distribution
and, possibly, a different thermal status of the wind.
%%%It is currently thought that the vast majority of ultraluminous X-ray sources (ULXs) 
%%%is powered by neutron stars and stellar-mass black holes accreting at rates 
%%%which may exceed the Eddington limit by factors of a few up to hundreds. 
%%%At these high accretion rates, radiation pressure is expected to inflate 
%%%the accretion disc and drive fast winds at significant fractions of the speed of light. 
%%%Evidence of such winds has been found in recent work with high-resolution grating  
%%%spectrometers (RGS) aboard XMM-\textit{Newton} as well as moderate-resolution detectors.
%%%The thick disc structure and the spectral energy distribution resembling those of a broadened 
%%%disc significantly differ from those of thin disc accretors such as sub-Eddington active galactic
%%%nuclei and X-ray binaries. 
%%%The thermal status of ULX winds, and of super-Eddington accretors
%%%in general, is therefore expected to depart from sub-Eddington accreting objects.
Here we show the first attempt to calculate the photoionization balance of the winds driven 
by strong radiation pressure in thick discs with a focus on ULXs {hosting black holes or 
non-magnetic neutron stars}. We find that the winds are generally in thermally stable
equilibrium, but long-term variations in the accretion rate and the inclination due to precession 
may have significant effects on the wind appearance and stability.
Our model trends can explain the observed correlation between the spectral residuals around 1 keV  
and the ULX spectral state.
We also find a possible correlation between the spectral hardness of the ULX,
the wind velocity and the ionization parameter in support of the general scenario.
\end{abstract}


\section{Introduction}
\label{sec:intro}

It is thought that when the Universe was young, black holes were accreting 
matter at impressive rates in order to build up ``fully grown'' supermassive black holes 
with masses of $\gtrsim10^9\,M_{\odot}$ in a few hundred million years 
(see, e.g., \citealt{Fan2003}, \citealt{Volonteri2003}).
At accretion rates approaching the Eddington limit, $\dot{M}_{\rm Edd}$, 
radiation pressure inflates the thin disc
producing a geometry similar to that of a funnel (see, e.g., \citealt{SS1973}) and
a wind is launched (see, e.g., \citealt{Poutanen2007}). 
These systems significantly differ from most common active galactic nuclei (AGN) 
and X-ray binaries (XRBs) which accrete at sub-Eddington rates 
(see, e.g., \citealt{Grimm2002,Aird2018}).
Winds driven by radiation pressure in a super-Eddington (or supercritical) regime 
can reach mildly-relativistic $\sim0.1c$ velocities (see, e.g., \citealt{Takeuchi2013}) 
and therefore may carry out a huge amount of matter and kinetic power, potentially 
altering both the accretion onto the black hole and the star-formation in the host galaxy 
(feedback, see, e.g., \citealt{Fabian2012} and references therein).

\subsection{Highly-accreting supermassive black holes}

Nowadays, it is still difficult to study the most distant ($z>6$) and rare quasars, 
but there is evidence that some nearby supermassive black holes are accreting at 
significant fractions of the Eddington limit or beyond such as those powering
Narrow Line Seyfert 1 galaxies (NLS1, e.g. \citealt{Boroson2002}). 
However, it is difficult to determine the bolometric luminosity in Eddington units 
of NLS1s as most disc emission is produced in the UV, significantly affected by  
interstellar absorption and several systematic effects by up to an order of magnitude, 
similarly to the highly uncertain mass estimates 
(see, e.g., \citealt{Vasudevan2007, Jin2012, CastelloMor2016, Buisson2018}).

Evidence of relativistic winds in NLS1 has indeed been found
(see, e.g., \citealt{Hagino2016, Kosec2018c}) and in some cases 
the strength of the absorption lines correlates with the variability of the ionizing continuum 
as expected for radiatively-driven winds (see, e.g., \citealt{Matzeu2017, Parker2017, Pinto2018}). 
However, these ``ultrafast outflows'' are very similar to those found in several 
AGN accreting at about the Eddington limit or 1-2 orders of magnitude below (see, e.g., 
\citealt{Pounds2003, Reeves2003, Tombesi2010}), which could be 
driven by magnetic fields (\citealt{Fukumura2017}) rather than radiation pressure 
(\citealt{King2015}). It is not easy to find and compare unique observable predictions 
of the theoretical wind models with the observations of winds. 
The study of the wind thermal stability can be a step forward in this field 
and, therefore, it is proposed in this paper.

\subsection{Ultraluminous X-ray sources}
\label{sec:ULX_intro}

The detailed study of accretion physics 
at extreme rates has been boosted in the last decade 
by extraordinary discoveries in the class of objects
known as Ultraluminous X-ray sources (ULXs), currently the best examples of 
compact objects surpassing the Eddington limit for long periods of time.
ULXs are bright, point-like, off-nucleus,
extragalactic sources with X-ray luminosities above $10^{39}$ erg/s,
resulting from accretion onto a compact object \citep{Kaaret2017}. 
Recent studies have shown that some ULXs are powered by
accretion onto neutron stars with strong magnetic fields 
($10^{9-14} G$, e.g. \citealt{Bachetti2014}, \citealt{Furst2016}, \citealt{Israel2017a}, \citealt{Israel2017b},
\citealt{Tsygankov2016}, \citealt{Carpano2018}, \citealt{Brightman2018}, 
\citealt{RodriguezCastillo2019}, \citealt{Sathyaprakash2019},
\citealt{Middleton2019}), which confirmed earlier speculation that ULXs 
are in majority stellar-mass compact objects $(<100 M_{\odot})$ at or in
excess of the Eddington limit (\citealt{King2001}, \citealt{Poutanen2007}, 
\citealt{Gladstone2009}, \citealt{Middleton2013}, \citealt{Liu2013}, \citealt{Motch2014}). 
A few ULXs might still host intermediate mass black holes
($10^{3-5} M_{\odot}$) at more sedate Eddington ratios 
(see, e.g., \citealt{Greene2007}, \citealt{Farrell2009}, \citealt{Webb2012}, \citealt{Mezcua2016}). 

\begin{figure}
  \includegraphics[width=1\columnwidth, angle=0]{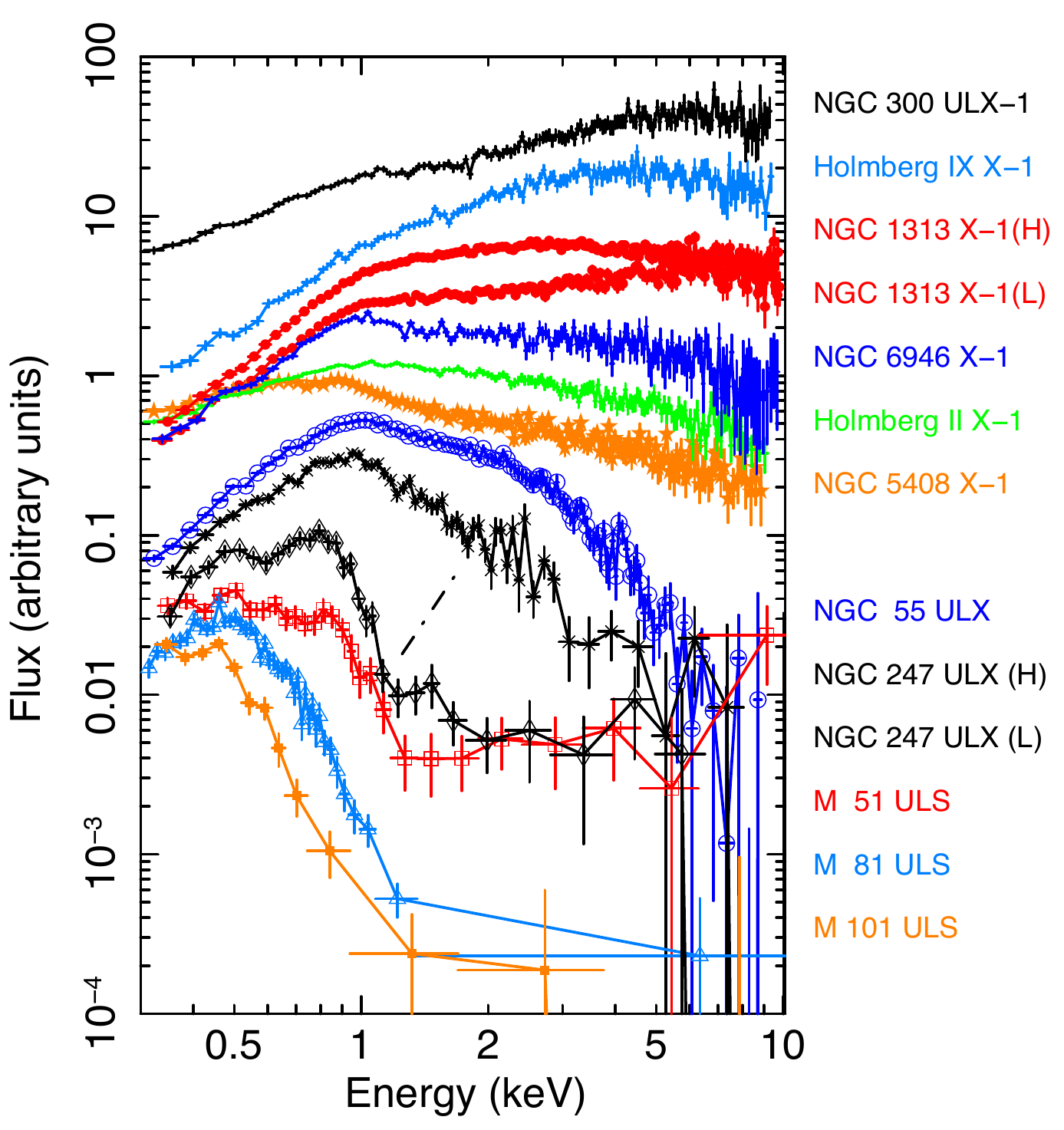}
  \vspace{-0.4cm} 
   \caption{X-ray spectra of some among the brightest ULXs with 
              hardness increasing from bottom to top. Y-axis units are 
              $E \times F_E$, but the spectra 
              were multiplied by constant factors - without altering the spectral shape - 
              for displaying purposes. For NGC 1313 and 247 we show two remarkably
              different spectral states.}
   \label{Fig:plot_ULX_sequence}
  \vspace{-0.4cm} 
\end{figure}

Before the discovery of ultraluminous pulsars (PULX) 
it was already thought that most ULXs were super-Eddington 
due to the presence of a strong turnover below $\sim7$\,keV 
in most ULX spectra (see, e.g., \citealt{Stobbart2006}, \citealt{Bachetti2013}).
In the softest states of ULXs {(e.g. in NGC 55 ULX)} 
the spectral curvature may start as low as 0.7 keV 
(see, e.g., \citealt{Gladstone2009}). 
It is difficult to explain the presence of this turnover in combination with the high
$10^{39-41}$\,erg/s luminosities invoking sub-Eddington accretion models.
ULXs represent a very complex category of objects with spectra of different
shapes and slopes. In Fig.\,\ref{Fig:plot_ULX_sequence} we show a hardness
sequence of ULX X-ray spectra uncorrected for interstellar absorption similar 
to \citet{Pintore2017} and \citet{Pinto2017}.
When modelled with a power-law, the spectra show the photon index ranging
from $\sim1$ for the hardest ULXs, to 2.5 for the soft ULXs
and above 4 for the ultraluminous supersoft sources (ULS or SSUL,
see, e.g., \citealt{Feng2016}, \citealt{Urquhart2016}).

In general, ULX broadband X-ray spectra require 2-3 emission components
(with the hardest component possibly related to the accretion column onto the poles
of a neutron star since it dominates the energies with the largest pulsed fractions 
in the known ULX pulsars when pulsations are detected; 
\citealt{Walton2018a}).
A broad blackbody-like emission with temperatures around 1\,keV
is likely produced by the supercritical thick disc itself, whose photon emission 
is highly distorted by hot electrons in the inner regions 
(up-scattering) and by down-scattering with cool electrons in the wind
(see, e.g., \citealt{Gladstone2009}, \citealt{Sutton2013} and \citealt{Middleton2015a}).
{In the case of highly-magnetized neutron stars with moderate accretion rates 
a substantial fraction of the emission around 1\,keV may come from near the neutron star
surface}.
A soft component with temperatures around 0.1\,keV most likely 
arises from upper regions of the outer disc, where a radiatively-driven wind 
is expected to be launched at accretion rates comparable to or 
higher than the Eddington limit 
(see, e.g., \citealt{Ohsuga2005, Poutanen2007, Takeuchi2013}). 
It is also possible that some portions of the wind, particularly phases of the outflow with much lower 
velocities, will be launched by thermal heating close to the Compton temperature 
(e.g., \citealt{Begelman1983, Done2018}).
{Outflows can also be launched by strong magnetic fields (see, e.g., \citealt{Romanova2015,
Romanova2018, Parfrey2017}).}

\subsubsection{The winds discovery and the unification scenario}

The first discovery of powerful winds possibly driven by radiation pressure in ULXs 
was achieved by the recent detection of blueshifted absorption lines in high-resolution 
soft X-ray spectra provided by the XMM-\textit{Newton} Reflection Grating Spectrometers
(RGS, \citealt{Pinto2016nature, Pinto2017, Kosec2018a}). 
Further confirmation was obtained in the Fe\,K hard X-ray energy band 
(\citealt{Walton2016a}) and, in particular, for the ultraluminous X-ray pulsar
NGC 300 ULX-1 (or NGC 300 PULX) both in the soft 
and the hard energy bands \citep{Kosec2018b}.

Several features were already seen, albeit spectrally unresolved,
in the low-to-moderate resolution CCD spectra
(see, e.g., \citealt{Stobbart2006}). The study of high-spatial-resolution 
\textit{Chandra} data \citep{Sutton2015} and timing variation
\citep{Middleton2015b} suggested that the residuals are mainly produced 
within a small region and are intrinsic to the ULX itself,
providing the first evidence of a wind (see also \citealt{Middleton2014}).

On the basis of this theoretical, computational and observational evidence, 
a unification scenario has been proposed according to which ULX spectral
properties depend on their accretion rates and inclination angles
(see, e.g., \citealt{Gladstone2009}, \citealt{Middleton2011}, \citealt{Middleton2015a}, 
\citealt{Sutton2013}) and on the ionization state
and complexity of the wind in the line of sight \citep{Pinto2017}.

\begin{figure}
  \includegraphics[width=1\columnwidth, angle=0]{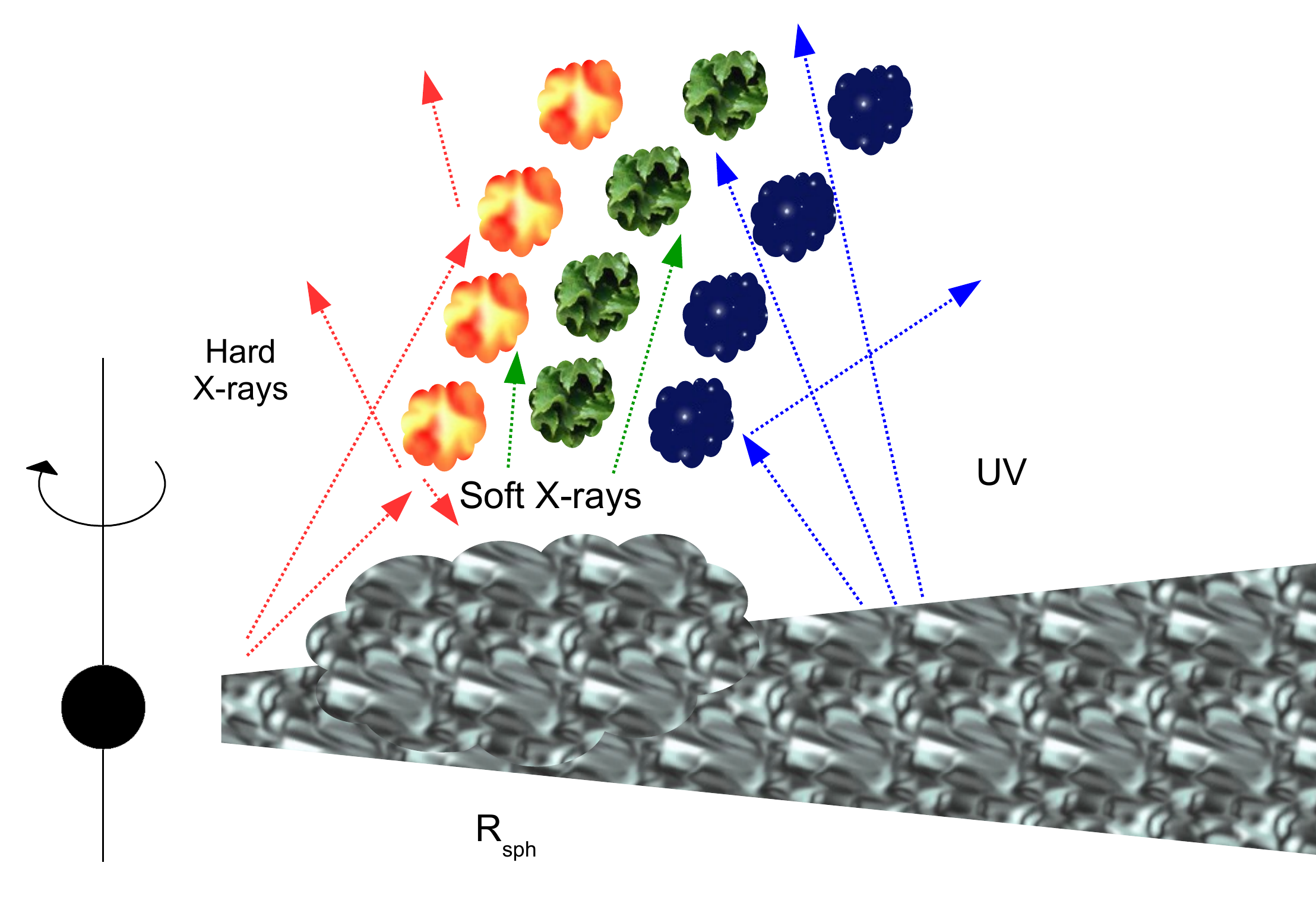}
  \vspace{-0.3cm}
   \caption{Schematic cartoon of a wind driven by radiation pressure in a 
                 super-Eddington accretion disc of a stellar mass black hole. 
                 The thickness of the disc and the high optical depth of the wind cause
                 the inner portion of the wind to be exposed to a harder X-ray continuum
                 compared to the outer regions, meaning that they are likely in a different thermal balance. }
   \label{Fig:plot_geometry}
  \vspace{-0.3cm}
\end{figure}

Fig.\,\ref{Fig:plot_geometry} shows an artistic impression of a wind
blowing from the supercritical region around the spherization radius,
$R_{\rm sph}$, of an accretion disc of a black hole.
In general, we should expect high ionization states of the wind at low
inclinations from the disc rotation axis (on-axis) 
and progressively lower ionization at higher inclinations.
However, the outflow is at least in part clumpy and optically thick, which implies 
that each line-of-sight (LOS) is dominated by a fraction of the wind
exposed to a certain ionizing field and therefore 
characterized by a different thermal equilibrium.

{A broadly similar scenario might apply for neutron star ULXs, provided that the magnetic field is not so strong that it truncates the disc before the thick inner regions form and launch the associated disc wind
(see, e.g., \citealt{Takahashi2017}).
\citet{Mushtukov2019a} demonstrated that this should be the case for magnetic fields of at least $\sim10^{12}$ G (and below). 
While there are cases which may have much stronger magnetic fields
(\citealt{Tsygankov2016, Brightman2018}), in general the field strengths in neutron star ULXs are still hotly debated
(see, e.g., \citealt{Middleton2019} and \citealt{KingLasota2019}), and there are also cases where there are indications that the field could be much more modest (e.g. \citealt{Walton2018b}). Indeed \citet{Kosec2018b}
report the detection of a powerful outflow in the ULX pulsar NGC 300 ULX-1, implying that the field strength and configuration permit these winds in at least some cases.}
{Nonetheless, the scheme in Fig.\,\ref{Fig:plot_geometry} is 
optimal for black holes and non-magnetic neutron stars.}

So far, no work has focused on the study of the thermal state
of the winds in ULXs and its effect on the variability and detectability
of the features imprinted on the ULX X-ray spectra. 
This is crucial in order to probe and understand whether the proposed
unification scenario actually works.
This is further complicated as three-dimensional radiation-hydrodynamic simulations 
of outflows from supercritical accretion have shown that the wind is  
clumpy and may vary on very short (seconds) time scales \citep{Kobayashi2018}. 
{Regardless of the exact launching mechanism (e.g. radiation pressure or 
magnetic pressure), it is still important to understand the thermal stability and other
physical properties of
ULX winds and, more generally, of winds observed in super-Eddington accretors.
ULX winds, for instance, might be relevant to understand the existence of superbubbles 
with sizes of about $\sim100s$\,pc around many ULXs (e.g., \citealt{Pakull2002,Pakull2010}).}

This paper is therefore dedicated to the study of the stability of ULX winds
and their comparison to winds found in better known X-ray sources 
such as Narrow Line Seyfert 1, classical sub-Eddington Seyfert 1 nuclei 
and classical supersoft sources such as white dwarfs burning hydrogen 
(some of which are novae).
This paper is structured as follows. 
We present the X-ray objects of this study in Sect.\,\ref{sec:sources}
and the computation of stability curves in Sect.\,\ref{sec:stability_curves}. 
We discuss the results and provide insights on future X-ray missions in 
Sect.\,\ref{sec:discussion}  
and give our conclusions in Sect.\,\ref{sec:conclusion}.
Some technical detail on our analysis is reported in Appendix\,\ref{sec:appendix}.
 
\begin{table*}
\renewcommand{\tabcolsep}{2mm}
\renewcommand{\arraystretch}{1.2}
%\footnotesize
\caption{\label{table:ULX} Characteristics of the brightest ULXs and XMM-\textit{Newton} exposure times} %%%%%% Maybe a column with the wind velocities?
\begin{tabular}{c|cccc|cccc|cc}
\hline
Source & d (Mpc)\,$^{(a)}$  & $L_{X}^{40}$ & Flux & $HR$ & Wind & $v_w$ & $\log \xi$ & t$_{\rm exp}^{\rm tot}$\,$^{(b)}$ & Similar sources   \\   %%% $\overline{t_{\rm exp}}$ (ks) $^{(b)}$
\hline                                              
{NGC 300 ULX-1}      & 2.0   & $0.5$ & $4$     & $0.7$  & D  & $0.24$  & $3.9$  & 212   & NGC 7793, NGC 5907, M 82 (PULXs) \\   %%%  &  $106$   
{Holmberg IX X-1} & 3.8   &  $4$   &  $15$  & $0.6$  & T  & $0.23$  & $4.5$  & 161   & IC 342 X-1, NGC 1313 X-2          \\    %%%  &  $18$     
{NGC 1313 X-1}    & 4.2   & $2$    & $5$     & $0.5$  & D  & $0.19$  & $2.3$  & 750   & NGC 5204 X-1                             \\    %%%  &  $107$   
{NGC 5204 X-1}    & 4.8   &  $1$   &  $2$    & $0.4$  & D  & $0.34$  & $3.0$  & 160   & NGC 1313 X-1                             \\    %%%  &  $27$     
{NGC 5408 X-1}    & 5.3   & $1$    & $3$     & $0.2$  & D  & $0.20$  & $2.1$  & 655   & NGC 6946 X-1, Holmberg II X-1  \\    %%%  &  $109$   
{NGC 55 X-1}        & 2.0   & $0.2$ & $3$     & $0.1$  & D  & $0.19$  & $3.3$  & 120   & NGC 247 X-1, M81 ULS              \\    %%%  &  $120$   
%%%{NGC 7793 P13} ?  & 3.6  & $1$ & $5$    & $0.3$  & T &  188      &  7  &  $26.9$     & NGC 300, 5907 X-1, M 82 X-2 \\
%%%{M 81 ULS}          & 3.6   & $??$ & $??$    & --  & 0        & 0   &  $0$       & $0.01$     & M 51-101 ULS \\ 
\hline
\end{tabular}
% \vspace{0.2cm}

{$^{(a)}$Distances are from the Nasa Extragalactic Database. 
               Unabsorbed X-ray $({0.3-10\, \rm keV})$ luminosities $L_{X}^{40}$ are in units of
               $(10^{40} \rm erg\,s^{-1})$, observed fluxes are in units of $10^{-12}$ erg s$^{-1}$ cm$^{-2}$ 
               and hardness ratios ($HR = L_{2-10 \rm keV} / L_{0.3-10 \rm keV}$)
               are estimated adopting a multi- component emission model for the XMM/EPIC-pn 
               spectra of the observations with evidence of winds 
               (see Fig.\,\ref{Fig:plot_ULX_sequence}). 
               The wind velocities are in units of speed of light
               and the ionization parameters $\log \xi$ are in units of erg/s cm.
               For more detail on the wind properties, see Table\,\ref{table:ULX_winds}.
               $^{(b)}$ XMM-\textit{Newton}/RGS total exposure time in ks. 
               Wind detections are labelled as $D$ (or $T$) if their significance is above (or below) $3\,\sigma$.
    {The X-ray luminosities refer to the time-average levels measured during the XMM-\textit{Newton}
               observations,
               we do not report their statistical uncertainties since they are much smaller than the flux variability
               (see, e.g., Fig. \ref{Fig:plot_wind_par}).
               Uncertainties on other parameters are reported in Table\,\ref{table:ULX_winds}.
               For more detail on the exposures used here see Sect. \ref{sec:appendix_obs}.}}
%\vspace{-0.5cm}
\end{table*}

\begin{table}
\caption{List of non-ULX sources used for comparison.}  
% \vspace{-0.2cm}
\label{table:noULX}      % is used to refer this table in the text
\renewcommand{\arraystretch}{1.2}
 \small\addtolength{\tabcolsep}{-2pt}
 
\scalebox{1}{%
\begin{tabular}{c c c c}     
\hline  
Source                   &   Source type            & $L_{\rm Edd}$ peak\,$^{(a)}$   & $M_{\odot}$       \\
\hline                                                                                                           
NGC\,5548            &  Seyfert 1 / AGN        &       0.05    &  $4\times10^7\,^{(b)}$    \\
IRAS\,13224          &  NLS1 / AGN             &       5          &  $1\times10^6\,^{(c)}$    \\
V2491\,Cygni       &  Nova / WD               &       1          &  $1.3\,^{(d)}$   \\
Blackbody              &  50 eV model            &       --          &   --            \\
\hline                
\end{tabular}}

$^{(a)}$ X-ray luminosity in units of Eddington limit (with systematic uncertainties
$\Delta L_{\rm Edd}\sim50$\,\%).
$^{(b)}$ \citet{Pancoast2014}, $^{(c)}$ \citet{Alston2018}, \citet{Pinto2018},
$^{(d)}$ \citet{Hachisu2009}.
\end{table}

\section[]{The sources}
\label{sec:sources}

In this section we report the objects of our study. We include archetypal ULXs
focusing on three cornerstone states (soft, intermediate, hard) represented by
NGC 5408 X-1, NGC 1313 X-1 and the pulsar NGC 300 ULX-1.
All three show evidence of outflows albeit at different statistical significance owing to the 
different exposure times (\citealt{Pinto2016nature}, \citealt{Kosec2018b}).
Some characteristics of these sources are shown in Table\,\ref{table:ULX}
along with other ULXs among the brightest ones and their highest significance
wind components.
These three sources have been observed with several satellites yielding broadband 
spectra covering the energy domain from IR to hard X-rays, which is necessary to
compute photoionization balance.
We also give examples of some other bright ULXs with similar spectral and timing behaviour.

On a broader context, we also want to compare ULX outflows to those found
in other sources, particularly those where radiation pressure may play an 
important role (Table\,\ref{table:noULX}). 
Therefore, we include in our study the supersoft source nova V2491 Cygni,
which exhibited a thick 3000 km/s outflowing envelope 40 days after its outburst
(see, e.g., \, \citealt{Ness2011} and \citealt{Pinto2012b}) and
the highly-accreting NLS1 IRAS 13224-3809 that shows an ultrafast $0.2c$ outflow
responding to continuum changes in agreement with theoretical models of radiation
pressure driven winds \citep{Parker2017,Pinto2018}.

We also test a blackbody spectrum with a temperature of 50\,eV to mimic the 
spectra of supersoft ultraluminous X-ray sources (see, e.g., \citealt{Urquhart2016})
and tidal disruption events (TDEs, see, e.g., \citealt{Leloudas2016}).
A TDE is the disruption of a star that occurs near the event horizon of a `small'
supermassive black hole, $M_{\rm BH} \sim 10^{5-7}\,M_{\odot}$.
They are believed to reach fallback luminosities $100\times$ Eddington for a short period
(e.g. \citealt{Wu2018}) and show evidence of radiation-pressure driven winds 
(see, e.g., \citealt{Miller2015}).

For useful diagnostics, we also incorporate the sub-Eddington Seyfert type 1 AGN 
NGC 5548 because the spectral code
that we use ({\scriptsize{SPEX}}, \citealt{kaastraspex}) 
has the photoionization balance calculation for this source as default \citep{Steenbrugge2005}
and was used in previous work on ULX winds 
(\citealt{Pinto2016nature}, \citealt{Pinto2017}, \citealt{Kosec2018a}, \citealt{Kosec2018b}).

We will first compare the ULX NGC 1313 X-1 
(where we found the first evidence of winds) to different types of non-ULX sources.
Then we will focus on the thermal stability of ULX winds according to their ionization
field. 

Here we do not consider different cases of obscured / unobscured AGN as a detailed
analysis was already performed by \citet{Mehdipour2016} including a description of
the systematics introduced by using various photoionization codes.
Testing different codes, they found variations of 10--30\,\% 
in the value of the ionization parameter of the winds and 
the optical depth of the absorption lines, which are difficult to detect with the current
X-ray telescopes.

\section{Thermal stability}
\label{sec:stability_curves}

The {\scriptsize{SPEX}} code\footnote{https://www.sron.nl/astrophysics-spex} 
is a powerful package
to compute ionization balance of high-energy astrophysical plasmas. 
In particular, the newly implemented \textit{pion} code is optimized to perform 
instantaneous calculation of ionization balance in the regime of photoionization, 
modelling simultaneously the input continuum and the emission/absorption lines 
produced by the ionized gas.

Photoionization equilibrium is parametrized with the well known relationship 
\begin{equation}\label{Eq:ionization_parameter}
\xi= \frac{L_{\rm ion}}{n_{\rm H}\,R^2}
\end{equation}
where $\xi$ is the ionization parameter (a measure of the number of photoionizing photons
per particle), $L_{\rm ion}$ the ionizing source luminosity (usually taken between 1 and 1000 Rydberg,
i.e. 13.6 eV and 13.6 keV), $n_{\rm H}$ the hydrogen density 
and $R$ the distance between the plasma and the ionizing source
(see, e.g., \citealt{Krolik1981}).
A crucial input to the photoionization calculation is therefore the ionizing continuum or
the broadband spectral energy distribution (SED).
The ionization balance also slightly depends on element abundances; here we adopt
the recommended proto-Solar abundances of \citet{Lodders2009}, 
which are also the default in {\scriptsize{SPEX}}.
 
\subsection{Spectral energy distribution}
\label{sec:SED}

\begin{figure}
  \includegraphics[width=1\columnwidth, angle=0]{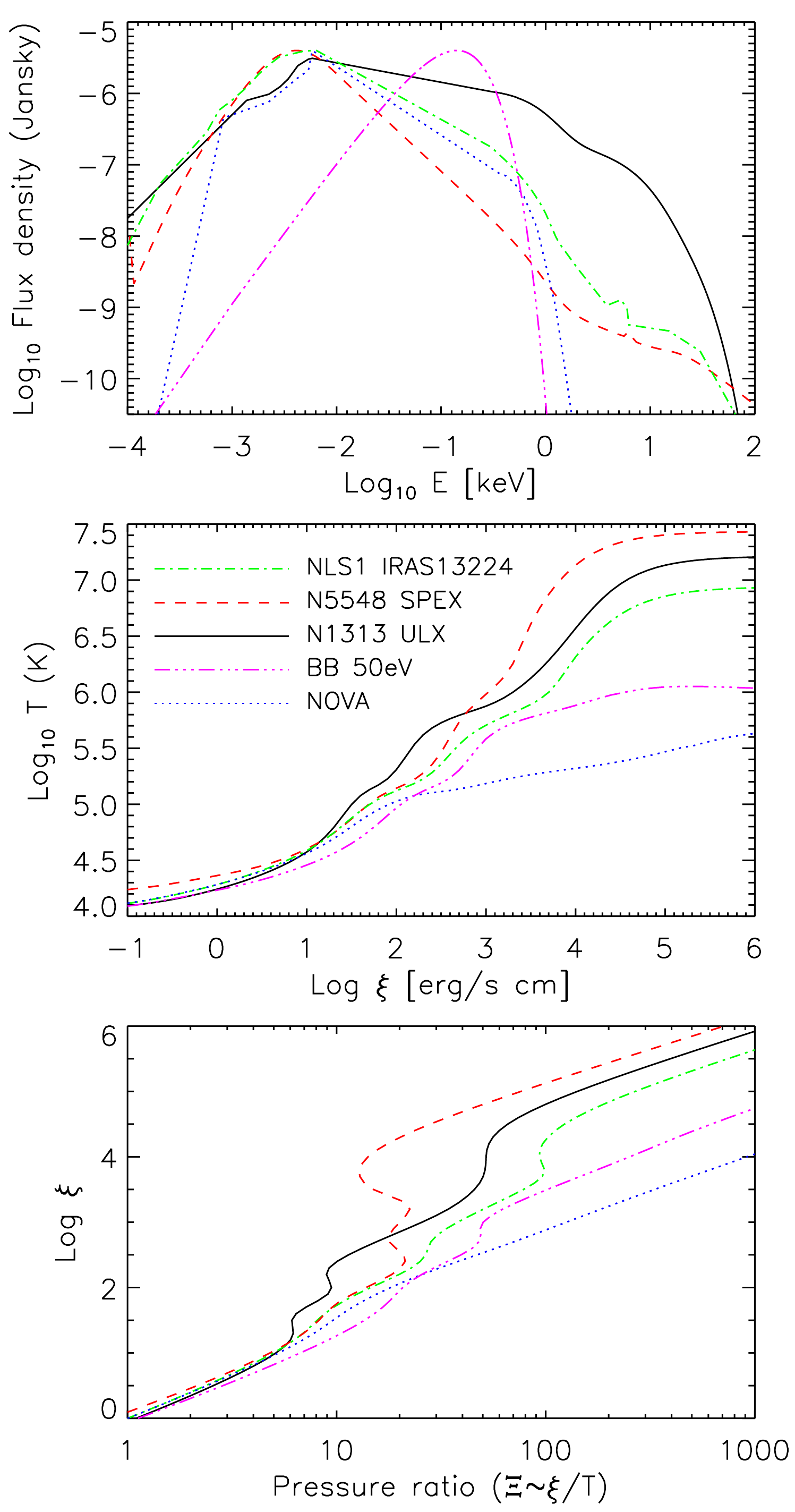}
   \caption{Spectral energy distribution (SED, top), $T-\xi$ (middle) and stability curves (bottom)
                 of ULX NGC 1313 X-1 compared to those of the AGN IRAS 13224 (NLS1) and 
                 NGC 5548 ({\scriptsize{SPEX}}), nova V2491 and a TDE-like 50 eV blackbody.
                 The SEDs are normalized, multiplying for a constant, for displaying purposes.}
   \label{Fig:plot_SED_Balance}
\end{figure}

In order to build the SEDs of our sources we extract fluxes through all the available 
data from optical to hard X-ray energies. In Fig.\,\ref{Fig:plot_SED_Balance} (top)
we compare the SED of ULX NGC 1313 X-1 with the SEDs of the other 
non-ULXs sources detailed in Table\,\ref{table:noULX}. 
All SEDs have been corrected for interstellar absorption and E(B-V) reddening.

Briefly, we take the time-averaged XMM-\textit{Newton} and NuSTAR X-ray spectra
of IRAS 13224-3809 from \citet{Parker2017} and the optical/UV data from the 
optical monitor (OM) time-average fluxes (\citealt{Buisson2018}).
For NGC\,5548 we use the default unobscured SED used by {\scriptsize{SPEX}} 
\citep{Steenbrugge2005}, while for nova V2491 Cygni we adopt the SED
produced in \citet{Pinto2012b}.

For NGC 1313 X-1 we adopt the time-average XMM-\textit{Newton}/EPIC spectrum 
(0.3--10 keV) where the wind 
was discovered \citep{Pinto2016nature} along with the hard X-ray fluxes provided
by NuSTAR (3--30 keV, \citealt{Bachetti2013}). Optical and UV fluxes are taken from
\citet{Yang2011}, whilst FUV fluxes are extracted from the 
XMM-\textit{Newton}/OM using a 4 arc seconds circle onto its X-ray centroid
and cross-checking with the source list produced 
by the XMM-SAS task {\scriptsize{OM\,CHAIN}}.
The effects due to the time variability of the SED and the optical/UV fluxes of
NGC 1313 X-1 are explored in 
Sect.\,\ref{sec:ngc1313_vs_variability}--\ref{sec:ngc1313_vs_screening}.

For the other two ULXs, namely NGC 5408 X-1 and NGC 300 ULX-1 we use the 
X-ray spectra from \citet{Pinto2016nature} and \citet{Kosec2018b}, respectively
(see also Table\,\ref{table:obs_log}). 
Optical and UV fluxes are obtained from HST measurements published in the literature
(\citealt{Grise2012}, \citealt{Lau2016}, \citealt{Villar2016}) 
and from the optical monitor on board 
XMM-\textit{Newton} where some literature data points are missing.
We prefer data from \textit{HST} observations which are deeper
and have higher spatial resolution than XMM-\textit{Newton}/OM.
Overall, the SEDs of the three ULXs look rather smooth and do not show
sharp jumps of flux when compared 
to the other sources with more accurate measurements outside the X-ray band 
like AGN and novae, see Fig.\,\ref{Fig:plot_SED_Balance} 
and \ref{Fig:plot_xiP_300_5408_1313}.
They resemble the broadened disc spectra in \citet{Sutton2013},
which suggests that the fluxes should be accurate with uncertainties less than an order
of magnitude, small enough to avoid dramatic uncertainties in the ionization
balance calculations.

The X-ray portions of the SEDs for the three ULXs under investigation
(NGC 1313 X-1, NGC 5408 X-1 and NGC 300 ULX-1) are determined 
by fitting the time-averaged XMM-\textit{Newton}/pn spectrum 
with a multi-component model consisting of blackbody, multicolour blackbody and
powerlaw as discussed in Sect.\,\ref{sec:intro} (see also \citealt{Kosec2018b}). 
The blackbody component ($T\sim0.15-0.25$\,keV) fits the disc upper photosphere,
the multicolour blackbody ($T\sim1-3$\,keV) reproduces the overall disc emission
({and/or any contribution from the boundary layer near a neutron star})
and the powerlaw ($\Gamma \sim 1.5-2.5$) accounts for emission from the 
inner disc (and eventual contribution from the accretion column onto a neutron 
star, see e.g. \citealt{Walton2018a}).
Neutral absorption from the Galactic interstellar medium and circumstellar medium
near the ULXs is reproduced by the $hot$ model in {\scriptsize{SPEX}} adopting
a low temperature $T=0.5$\,eV (see, e.g., \citealt{Pinto2013} and references therein).

{In summary, we have obtained the UV-to-IR fluxes at 2120, 2310, 2910, 3600,
4300 and 5500 and 9000\,{\AA} from the literature or the XMM-\textit{Newton}/OM,
the X-ray fluxes between 0.3--10\,keV ($\sim1.2-40$ \AA) from the XMM-\textit{Newton} 
RGS+pn spectral fits. The SED portions above 10 keV are either provided by NuSTAR
or extrapolated from our XMM-\textit{Newton} spectral model.
The SED range covering the FUV domain is missing due to interstellar absorption and
is therefore interpolated. This adds uncertainty in the photoionization calculation but we
stress that our SED shapes agree with those calculated for irradiated discs of black holes
in the regime of super-Eddington accretion \citep{Ambrosi2018}. 
We investigate some systematic effects due to uncertainties in the low-energy flux
in Sect. \ref{sec:ngc1313_vs_screening}.}

\subsection{NGC 1313 ULX and non-ULX objects}
\label{sec:ngc1313_vs_noULX}

We compute the photoionization balance running the {\scriptsize{SPEX}}
\textit{pion} code version 3.05.00 onto all SEDs.
In Fig.\,\ref{Fig:plot_SED_Balance} (middle panel) we show the detail of the 
photoionization balance computation (i.e. the temperature $T$ -- ionization parameter
$\xi$ curves). Some differences can be distinguished between 
these curves such as the variable slopes at intermediate values of the
ionization parameter ($\log\,\xi \sim 2-3$). 
The detail of how heating and cooling rates are computed for 
ULX NGC 1313 X-1 and, in particular, their trends with the ionization parameter
can be found in Appendix\,\ref{sec:appendix_rates}.
 
However, a more informative diagram is provided by the stability curves. Those can
be produced by computing the ratio between the radiation pressure ($F/c$) and 
the thermal pressure ($n_{\rm H}kT$), 
which is given by $\Xi = F / n_{\rm H} c kT = 19222 \, \xi / T$,
with $F = L_{\rm ion} / 4 \pi r^2$ \citep{Krolik1981}.
The stability (or $S$) curves computed for our sample of sources are shown
in Fig.\,\ref{Fig:plot_SED_Balance} (bottom panel). 
The $S$ curves are powerful diagnosis tools because along them 
heating equals cooling and, therefore, the gas is in thermal balance.

On the left side of the $S$ curve cooling dominates over heating, 
whilst on the right side, heating dominates over cooling. 
Most importantly, where the $S$ curve has a positive gradient, 
the photoionized gas is thermally stable or, in other words, small perturbations 
upwards in temperature will be balanced by an increase in cooling.
Similarly, small perturbations downwards will be balanced by increase in heating. 
Instead, in the points on the $S$ curve with a negative gradient, the plasma is thermally 
unstable and therefore any perturbation upwards (downwards) will cause further 
increase (decrease) of temperature with a dramatic change, if not loss, of the equilibrium.
In principle, any two components on the $S$-curve with the same $\Xi$ value are expected
to be in pressure equilibrium. 

The stability curves computed for the different sources (Fig.\,\ref{Fig:plot_SED_Balance})
show notable differences, with the harder sources exhibiting several unstable branches. 
Here with hard sources we actually mean `X-ray hard sources' i.e. those sources which 
have a high ratio between the fluxes calculated in the 2--10\,keV and 0.3--2\,keV energy bands,
respectively.
High-energy photons from the Fe K energy band and beyond significantly affect the
ionization balance by increasing the size of the unstable region
where, in principle, no plasma should be found
(see also \citealt{Krolik1981}).
On the other hand, the softest sources like novae or SSUL have rather stable curves.
Moreover, the high-temperature turnovers differ by up to two orders of magnitude
(Fig.\,\ref{Fig:plot_SED_Balance}, middle panel),
which indicates great differences in the Compton temperature. 
Such variations are expected in objects of different accretion rates and are relevant for
thermal and thermal-radiative winds since they can be suppressed if the Compton 
temperatures are too low (\citealt{Done2018}).

$S$ curves of AGN (and classical X-ray binaries) have been studied in great detail
and their instability branches are ubiquitously found 
(see, e.g., \citealt{Krolik1981, Reynolds1995, Netzer2003, Steenbrugge2005,
Chakravorty2016, Mehdipour2016, Higginbottom2017}).
Of course, owing to its SED, the stability curve of archetypal ULX NGC 1313 X-1 
fits in between those of AGN and soft sources, showing only some small branches
of instability.

\subsection{NGC 1313 ULX and other ULXs}
\label{sec:ngc1313_vs_ULX}

A comparison between the $S$ curves of the intermediate-state 
ULX NGC 1313 X-1 with harder (Seyfert 1 AGN) and softer (supersoft source, SSS, nova) 
spectra provides a crude idea of how thermal instability increases for the wind parts 
which are subjected to harder SEDs. 
There is currently a strong likelihood that ULXs are characterized
by geometrically-thick discs and optically-thick outflows, at least for ULXs powered by
black holes or neutron stars with limited magnetic fields (e.g. $\lesssim10^{12} G$). 
It is therefore thought that the spectral hardness of a ULX is 
due to a combination of accretion rate and inclination. In fact,
at lower inclinations (face on) we should see harder spectra
(e.g., \citealt{Middleton2015a, Feng2016, Urquhart2016, Pinto2017}). 

It is very likely that the inner portions of the wind are exposed to a harder photoionizing field
compared to the outer regions or, in other words, the inner regions may see beamed hard X-rays, 
while the radiation field in the outer regions may be dominated by UV and soft X-rays (see 
Fig.\,\ref{Fig:plot_geometry}). If this scenario applies then the gas phases in these regions 
will have radically different stability curves.

\begin{figure}
  \includegraphics[width=1\columnwidth, angle=0]{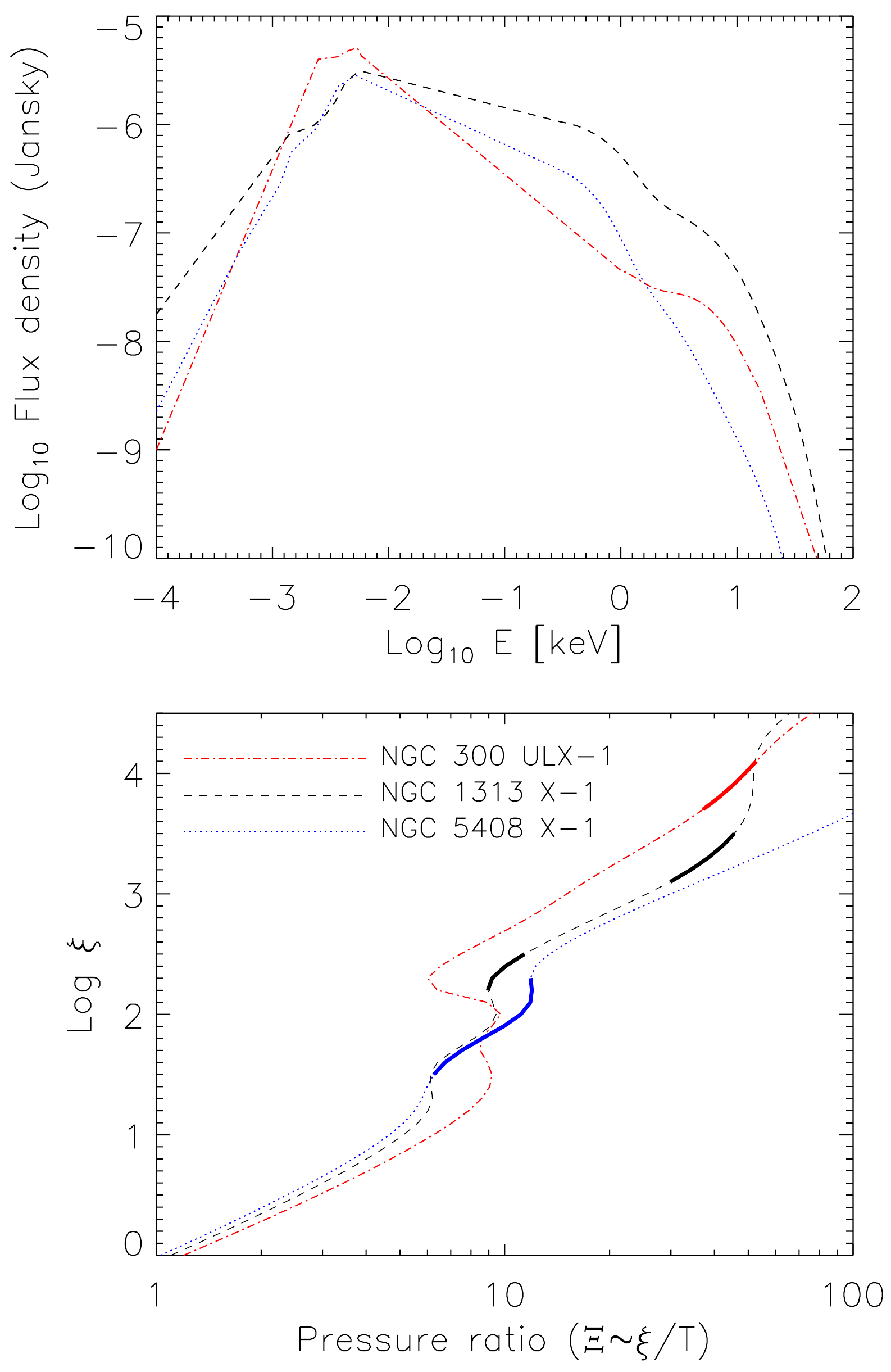}
  \vspace{-0.6cm} 
   \caption{SEDs of three ULXs of different spectral hardness 
                 and the corresponding stability $S$ curves. 
                 Thick regions on the $S$ curves show the wind detections 
                 for these sources (\citealt{Pinto2016nature, Walton2016a, Kosec2018b}).
                 The SEDs have been normalized for displaying purposes
                 similarly to Fig.\,\ref{Fig:plot_SED_Balance}.}
   \label{Fig:plot_xiP_300_5408_1313}
  \vspace{-0.3cm} 
\end{figure}

In Fig.\,\ref{Fig:plot_xiP_300_5408_1313} (top) we compare the SEDs of three ULXs 
with different X-ray spectral shapes: 
NGC 300 ULX-1 (hard), NGC 1313 X-1 (intermediate) and 
NGC 5408 X-1 (soft, see also Table\,\ref{table:ULX}).
We have computed the ionization balance for the other two ULXs with the same code
used for NGC 1313 X-1 in Sect.\,\ref{sec:ngc1313_vs_noULX}
and found significant differences in their stability curves 
(see Fig.\,\ref{Fig:plot_xiP_300_5408_1313}, bottom).
As expected, the harder ULX has longer branches of instability.
This means that, according to the ULX unification scenario, there should be a
systematic increase of instability for wind portions towards the inner regions of 
the accretion disc.

In Fig.\,\ref{Fig:plot_xiP_300_5408_1313} (bottom) we also indicate 
the ionization parameters and the corresponding pressure ratios
of the winds detected in these ULXs on each $S$ curve.
For NGC 1313 X-1, there are two regions 
due to two alternative solutions for the Fe L and Fe K energy bands
(\citealt{Pinto2016nature, Walton2016a}, respectively).
It appears that the wind ionization state increases with the spectral
hardness of the source (and likely decreases with the viewing angle) and that the measurements
are formally acceptable as none falls along instability branches.
For NGC 5408 X-1, instead, the outflow has a more complex structure 
with two components of comparable ionization state ($\log \xi \sim 1.5-1.9$ 
and $1.7-2.1$) but different velocities (\citealt{Pinto2016nature};
see also Table\,\ref{table:ULX_winds}).

\subsection{NGC 1313 ULX: Spectral Variability}
\label{sec:ngc1313_vs_variability}

We normally consider as classical ULXs those sources which surpass the Eddington limit 
of a 10 Solar mass black hole for long periods of time. However, these 
persistent beacons can show remarkable X-ray variability on time scales from hours
to several days. PULXs for instance show long-term variability associated to
either precession (see, e.g., \citealt{Middleton2018}) or 
switching between two different regimes that may be interpreted 
as a propeller phase (see, e.g., \citealt{Tsygankov2016}). 
Luminosities of ULXs can vary by an order of magnitude, followed by changes in the spectral 
slope by 20\% or more (see, e.g., \citealt{Kajava2009}).
It is therefore a useful exercise to compute the ionization balance for ULXs where 
the spectral shape significantly changes with time in order to probe any effects 
on the wind stability (or on the portion of the wind in our line-of-sight).

Of course, NGC 1313 X-1 provides an excellent workbench thanks to the dozens of 
dedicated observations taken with XMM-\textit{Newton} since early 2000s and 
owing to its high spectral variability (see, e.g., \citealt{Middleton2015b}).
We therefore focus on three characteristic spectra of this ULX which show that the 
source alternates between soft-ultraluminous, hard-ultraluminous and 
bright-broadened-disc states, respectively (according to the classification of
\citealt{Sutton2013}; for more detail see Table\,\ref{table:obs_log}).

\begin{figure}
  \includegraphics[width=1\columnwidth, angle=0]{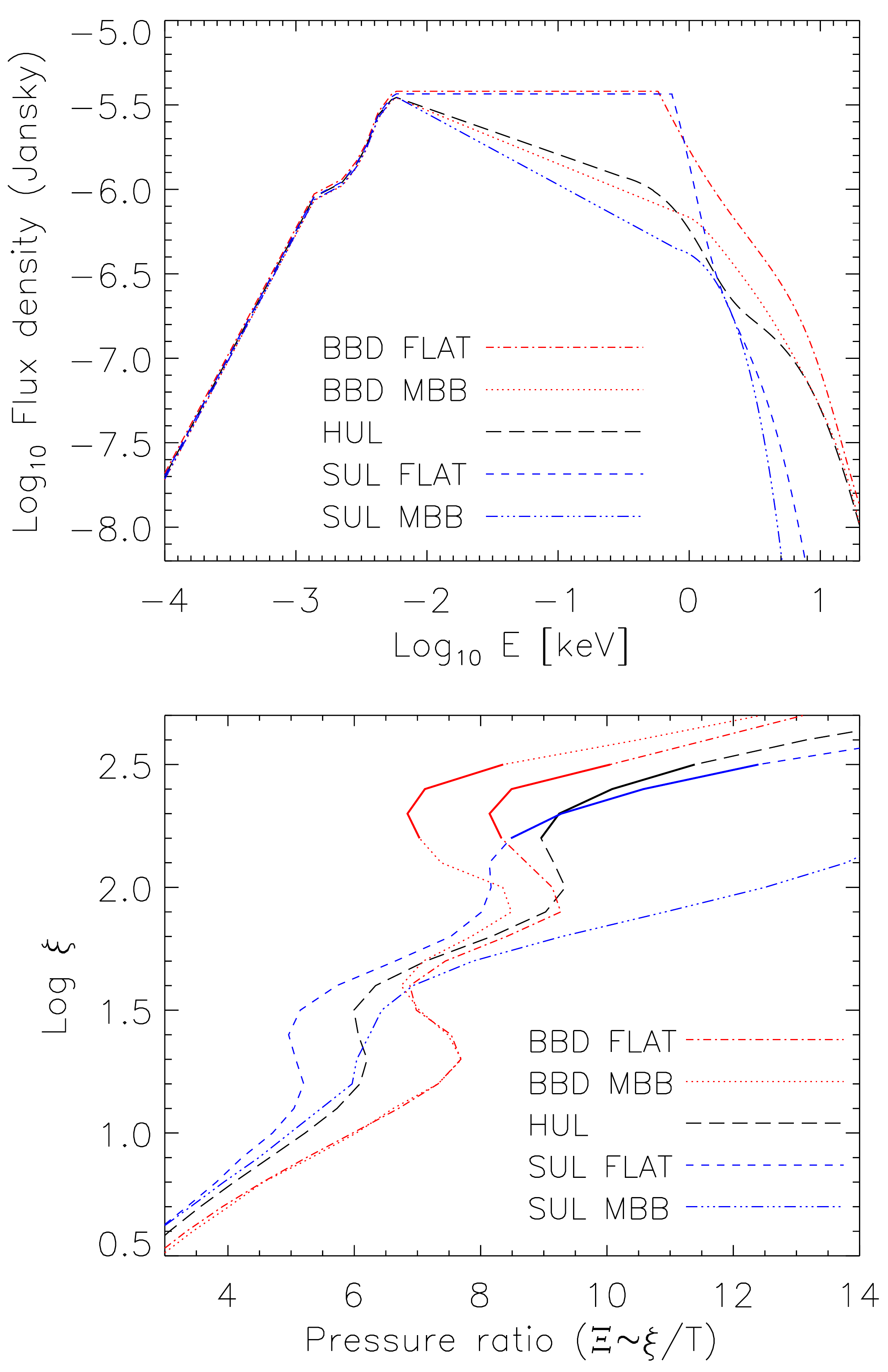}
  \vspace{-0.6cm} 
   \caption{Spectral energy distribution (top) and stability curves (bottom) of ULX NGC 1313 X-1
                 calculated for different states (soft, SUL, bright-broadened-disc, BBD 
                 and hard ultraluminous, HUL; see Table\,\ref{table:obs_log}).
                 The solid intervals in the bottom plot indicate the RGS best fit wind
                 solution ($0.2c$) obtained in \citet{Pinto2016nature}. 
                 Two different continuum models for the BBD and 
                 SUL SEDs are tested here (see Sect.\,\ref{sec:ngc1313_vs_variability}).
                 During the brightest BBD state the gas might move 
                 towards an unstable branch.}
   \label{Fig:plot_SED_Balance_ngc1313_variability}
  \vspace{-0.3cm} 
\end{figure}

NGC 1313 X-1 normally spends most of the time in the hard-ultraluminous state
with intermediate brightness compared to the soft (X-ray faint) and the 
broadened-disc (X-ray bright) states.
The corresponding SEDs are shown in  
Fig.\,\ref{Fig:plot_SED_Balance_ngc1313_variability} (top). 
For simplicity, we adopt the same values of optical and UV flux as given by
the time-averaged SED (see Sect. \ref{sec:ngc1313_vs_screening}
for a discussion on effects due to variable optical/UV fluxes). 
These broadband spectra represent a small sample,
but they provide a good indication of how much the ionization field may vary
and, therefore, how different will be the SED seen by the outflowing gas
depending on its location (and our line of sight).

The standard continuum model consisting of absorbed powerlaw, blackbody and 
modified blackbody overestimates the flux below 1 keV due to the powerlaw
steep index ($\Gamma\sim2$). This effect is stronger for the X-ray bright and faint
spectra (BBD and SUL, see Fig.\,\ref{Fig:plot_SED_Balance_ngc1313_variability} top)
due to their strong spectral curvature and requirement of high column density
(an order of magnitude above the interstellar value, 
$N_{\rm H, ISM}\sim5\times10^{20}$\,cm$^{-2}$). 
For this reason, we cut their extrapolation 
below 1 keV forcing a flat plateau (see the blue dashed and red dashed-dotted lines).
Alternatively, we also test another model where the powerlaw is substituted by another
modified blackbody, i.e. $hot*(bb+mbb+mbb)$. 
This model reproduces the spectral 
curvature without the need of high neutral absorption 
(see the blue dashed-dotted and red dotted lines) and indicates the order of 
magnitude of uncertainty in the (unabsorbed) SEDs of these states.
The intermediate, hard-ultraluminous state is not strongly affected by this issue
and, therefore, we just keep the results obtained with the standard model.

We compute the ionization balance for these SEDs of NGC 1313 X-1 as done
for the time-average spectrum and show the stability curves in 
Fig.\,\ref{Fig:plot_SED_Balance_ngc1313_variability} (bottom).
As expected, the $S$ curves significantly differ; in particular the broadened-disc
SED exhibits a large instability branch for ionization parameters $\log \xi \sim 1.8-2.3$,
which is smaller in the intermediate-hard state. The soft state shows instead
a generally stable curve very similar to soft ULXs, novae and other SSS
(see Fig.\,\ref{Fig:plot_SED_Balance} and \ref{Fig:plot_xiP_300_5408_1313}).
The uncertainties in the SEDs due to the soft X-ray extrapolation do not seem
to have drastic effects on the thermal stability of the wind.

\subsection{NGC 1313 ULX and optical/UV screening}
\label{sec:ngc1313_vs_screening}

It is difficult to build a proper spectral energy distribution without simultaneous
observations in all energy domains going from the optical to the X-rays. 
In most cases quasi-simultaneity with optical/UV data taken at slightly different time
from the X-ray can provide reasonable measurements as we might expect the optical/UV
not to vary on time scales as short as X-rays owing to the larger emitting region. 
In ULXs, of course, the scenario is complex as the bulk of the UV (and possibly the optical) 
emission may come from the outer disc itself but some contamination
can be expected from the companion star (e.g., \citealt{Grise2012, Ambrosi2018}). 
\citet{Sutton2014} have shown that most objects at $\sim10^{39}$ erg/s exhibit  
disc reprocessing fractions similar to sub-Eddington objects, but at higher luminosities
X-ray reprocessing seems to increase, possibly due to scattering of wind X-rays 
back to the outer disc. 

It is important to notice that the full optical-UV-X-ray SED that 
we observe might not actually be 
representative of the SED that ionizes all the different regions of the wind.
The thick inner disc and even the wind itself may self-screen the inner regions 
of the wind from some of the optical/UV emission arising from the outer disc
and the companion star. 

Fig.\,\ref{Fig:plot_geometry} shows a simplified, qualitative, description of 
a radiation-driven wind from a super-Eddington accretion disc
around a stellar-mass black hole or a neutron star.
The temperature at the spherization radius can be roughly expressed as
$T_{\rm sph}=1.5\,m^{-1/4}\,\dot{m}_0^{-1/2} \,(1+0.3\,\dot{m}_0^{-3/4})\, \rm keV$
(see, e.g., \citealt{Poutanen2007}),
which is of the order of 0.2\,keV for a black hole with a mass 
$M=m\,M_{\odot}=10\,M_{\odot}$
and an accretion rate $\dot{M}=15\,\dot{M}_{\rm Edd}=\dot{m}_0\,\dot{M}_{\rm Edd}$
{(where $\dot{m}_0 = \dot{m}$ at $R>R_{\rm sph}$ and we adopt the common definition 
$\dot{M}_{\rm Edd}=L_{\rm Edd}/\epsilon c^2=1.86\cdot10^{18}(M/M_{\odot}) {\rm \, g \, s}^{-1}$
with a radiative efficiency $\epsilon=8$\%)}. 
A comparable temperature is found in the soft X-ray component 
of almost all ULXs.
Most X-rays will be produced within $T_{\rm sph}$, while at larger radii
the flux will peak in the UV and optical energy bands.

\begin{figure}
  \includegraphics[width=0.985\columnwidth, angle=0]{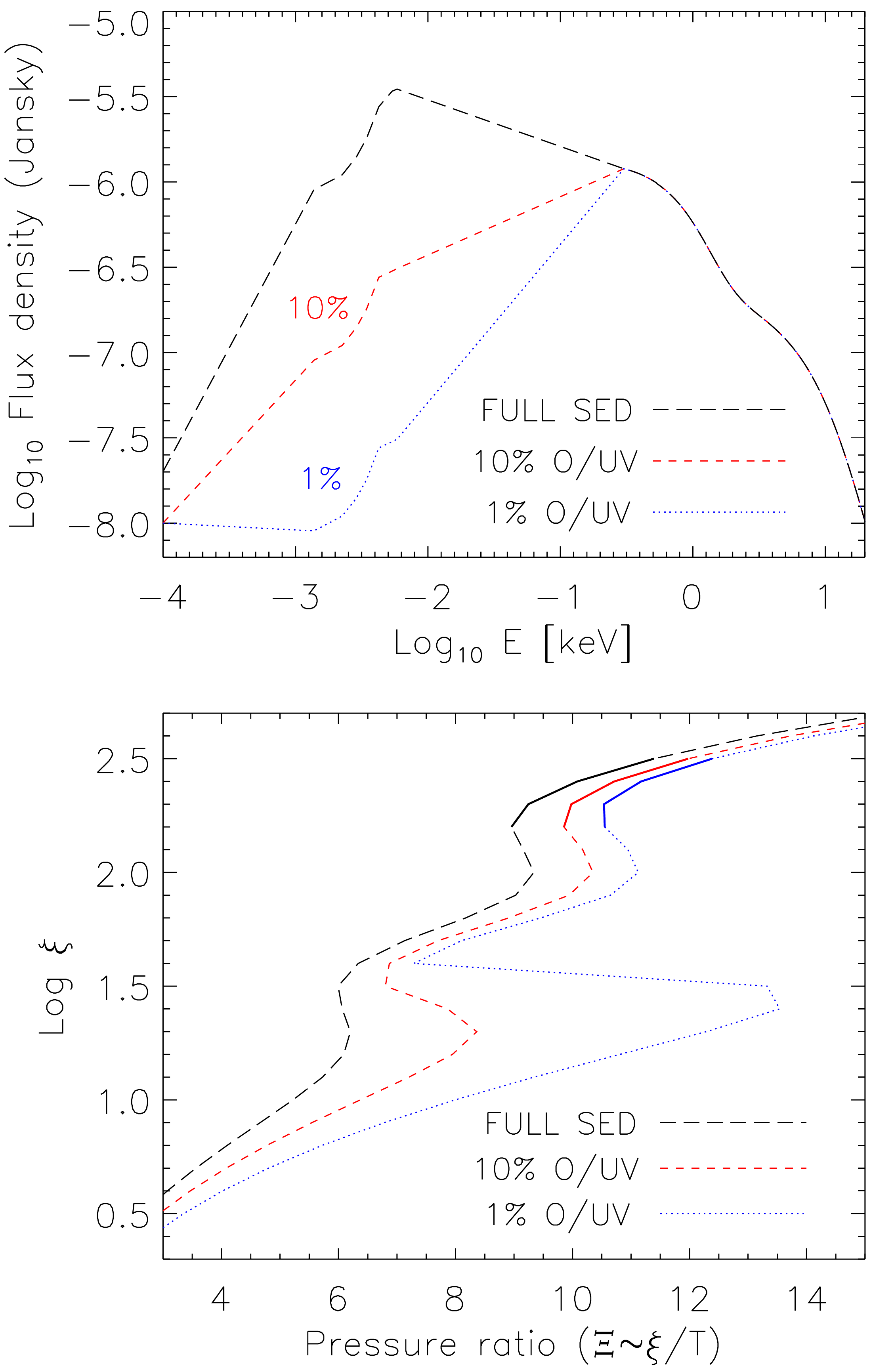}
  \vspace{-0.3cm} 
   \caption{Spectral energy distribution (top) and stability curves (bottom) of ULX NGC 1313 X-1
                 calculated for the hard ultraluminous state, HUL (see Table\,\ref{table:obs_log}),
                 as compared to those in the cases where either 10\% or 1\% of the 
                 optical/UV flux reaches the absorbing gas.
                 The solid intervals in the bottom plot indicate the RGS best fit wind
                 solution ($0.2c$) obtained in \citet{Pinto2016nature}.
                 At high optical/UV screening, some portions of the wind could be unstable.}
   \label{Fig:plot_SED_Balance_ngc1313_screening}
  \vspace{-0.3cm} 
\end{figure}

The uncertainties in the SED shape and, particularly, in the UV/optical irradiation of
the winds can be tested by computing the ionization balance
for SEDs where the optical/UV fluxes are 1-2 orders of magnitude below the threshold
measured with the XMM-\textit{Newton}/OM or other facilities. 
We therefore construct ad-hoc SEDs with low-energy attenuation
where only 1\% or 10\% of the optical/UV fluxes are transmitted and therefore received
by the plasma in our LOS (see Fig.\,\ref{Fig:plot_SED_Balance_ngc1313_screening}, top).
This is a very simplified way to simulate partial covering of outer UV/optical photons 
by inner optically thick clouds.
We use as template the SED of NGC 1313 X-1 during the most common hard-ultraluminous state
where a wind was detected. 

The stability curves show a progressive extension of the unstable branch for optical/UV 
transmission decreasing from 100\% to 10\% and then 1\% 
(see Fig.\,\ref{Fig:plot_SED_Balance_ngc1313_screening}, bottom).
The effects of optical/UV screening is however more pronounced at low ionization
parameters ($\log \xi \lesssim1.5$) which might not affect the current wind detection
($\log \xi \gtrsim2.2$).

\section{Discussion}
\label{sec:discussion}

The most common AGN rarely reach the Eddington limit. 
Other extreme objects such as the tidal disruption events 
may reach fallback luminosities $100\times$ Eddington for a short period 
(e.g. \citealt{Wu2018}), but they are rare and difficult to observe
due to their transient nature. 
It is also challenging to construct robust broadband SEDs for many AGN 
and TDEs because of extinction from neutral hydrogen, 
which mainly affects the energy range containing the SED peaks.

ULXs instead are X-ray bright and persistent, surpassing the Eddington limit
for long periods of time. The search and study of radiatively driven winds 
in these sources may therefore provide clues on the way primordial black holes 
grew up when the Universe was a few hundred million years old. 
%%%So far, no work has investigated the thermal stability of ULX winds and, therefore, 
In this paper we provide a first attempt to understand the thermal state of ULX winds,
their relationship with the LOS (that can change due to precession,
e.g., \citealt{Middleton2018})
and the accretion rate (e.g., \citealt{Poutanen2007, Middleton2015a}). 

\subsection{Thermal stability of winds in ULXs}

%%%Ultrafast outflows in ULXs are a new field of research. Some important,
%%%early, indications of ULX winds were given by unresolved soft X-ray spectral 
%%%features in high-quality CCD spectra (e.g., \citealt{Middleton2014, 
%%%Middleton2015b}).
%%%This new channel of research has been made possible through the recent discoveries 
%%%of outflows in the ULXs with the deepest XMM-\textit{Newton} observations
%%%(\citealt{Pinto2016nature, Pinto2017, Walton2016a, Kosec2018a, Kosec2018b}).
%%%
%%%So far, no work has investigated the thermal stability of ULX winds and, therefore, 
%%%this paper focuses on calculating stability curves, comparing them
%%%to those better studied in AGN and checking the thermal stability of ULX winds 
%%%according to their observables.

In Sect.\,\ref{sec:SED} we have shown
broadband SED spectra of different types of sources (AGN, ULX, nova, etc). 
Then we have computed the ionization balance of plasma around
ULXs for different spectral states
and shown the stability curves, indicating combinations of $\xi-\Xi$
where the plasma is stable (Fig.\,\ref{Fig:plot_xiP_300_5408_1313}). 
The winds detected in the three ULXs under investigation 
(i.e. NGC 300 ULX-1, NGC 1313 X-1 and NGC 5408 X-1)
show $\xi-\Xi$ pairs corresponding to positive slopes of the $S$ curve, 
which suggests that they are in stable equilibrium
(as it would be required for them to be detectable).

It is important to notice that our results predict the existence of an 
{evacuated funnel, similar} to that one previously invoked in ULXs (e.g. \citealt{Middleton2015a}). 
In the innermost regions near the compact object, 
the wind likely becomes highly-ionized and/or thermally unstable, possibly 
providing a better view on the hard X-ray emitting region
{(see also Sec. \ref{sec:effects_from_incl_rate} for evidence from observations)}.
{Moreover, density variations might cause thermal instabilities and multiphase gas
(see, e.g., \citealt{Owocki2006}) as suggested by the relatively mild ionization parameters 
and complex wind structure found in some ULXs.}

\subsection{ULX winds in a broader context}

\subsubsection{Comparison with winds from AGN and XRBs} 

There have been several studies dedicated to outflows in AGN and 
classical X-ray binaries, as well as their thermal stability in the past decades. 
Winds are better characterized in these objects owing to their high brightness
due to either proximity (Galactic X-ray binaries) or huge luminosity (AGN).
Below we report a brief summary on such winds.

AGN show a variety of winds detected in the UV such as the $0-5000$\,km/s 
warm absorbers (see, e.g., \citealt{Reynolds1995,Steenbrugge2005,Laha2014}), 
the UV/X-ray obscurers that create temporary occultations and have comparable 
velocities (see, e.g., \citealt{Kaastra2014, Mehdipour2017, Turner2018}), 
the $\sim10\,000$\,km/s broad-absorption-line quasars (BALQSO, e.g., 
\citealt{Proga2000, Hewett2003, Dai2008, Leighly2014})
and the most extreme $0.1-0.3c$ ultrafast outflows detected in the X-ray band 
(see, e.g., \citealt{Pounds2003, Reeves2003, Tombesi2010}).
The highly ionized ultrafast outflows tend to occupy large branches of stability 
through the $S$ curves (see, e.g., \citealt{Danehkar2018,Kraemer2018}), 
while the warm absorbers and the other winds with intermediate ionization states
can fall near the regions of thermal instability 
($\log \xi\sim1-2$, 
see e.g. \citealt{Krolik1981, Krongold2003, Netzer2003, Detmers11}).
The ultrafast outflows found in AGN are likely the most relevant comparison 
for the winds seen in ULXs as their similar properties ($\xi$, $v_{\rm outflow}, 
N_{\rm H}$) suggest some analogies in their launching mechanisms.

X-ray binaries often show highly ionized winds detected primarily in the Fe K band
(see, e.g., \citealt{Miller2004, Miller2006, Neilsen2009, Ponti2012}). These winds are typically
slow (a few 100s km/s, therefore associated with the disc); they are
often detected in the soft states and disappear when the source is in a hard state,
most likely due to the instability branch that stretches further at harder ionizing SEDs 
(see, e.g., \citealt{Chakravorty2016, Higginbottom2017}).
However, an acceleration of the flow or the exhaustion of the plasma during the soft state
are possible, viable solutions (see, e.g., \citealt{Gatuzz2019}).

SEDs of ULXs can significantly differ the one from the other,
but they often show shapes resembling that of a broadened disc blackbody or 
of a complex, super-Eddington, accretion disc 
(see, e.g., \citealt{Gladstone2009, Sutton2013, Middleton2015a}). 
Their stability curves are halfway between those of 
high-Eddington NLS1 and supersoft novae, having only small branches
of instabilities (see Fig.\,\ref{Fig:plot_SED_Balance}). 
The hard state ULXs such as the ultraluminous pulsar NGC 300 ULX-1 have in proportion
a larger fraction of hard X-ray photons which further extend the unstable section
(see Fig.\,\ref{Fig:plot_xiP_300_5408_1313}).
For PULXs with stronger magnetic fields, we can naturally expect more significant 
truncation of the inner accretion disc and strong polar inflow (at a given $\dot{M}$).
This not only decreases radiation pressure in the inner regions of the disc, but can 
also produce a harder SED, followed by increasing instabilities in the inner portions
of the wind. In other words, strongly-magnetized PULX will likely have weaker winds
(i.e. with lower outflow rates) than weakly-magnetized PULXs 
for the same values of $\dot{M}$.

{As mentioned before, outflows can also be launched by mild magnetic fields 
(see, e.g., \citealt{Romanova2015, Romanova2018, Parfrey2017}). A study of the launching
mechanism is however beyond the scope of our paper.}

\subsubsection{Effects of ULX winds on their surrounding}

This study of ULX winds at small scales might be relevant to understand
their effects at larger scales such as their role in the existence of ULX superbubbles 
with sizes of about $\sim100s$\,pc that are thought to be similar to the (radio) W\,50 nebula 
around SS\,433 (e.g., \citealt{Pakull2002,Pakull2010}).
For instance, \citet{Pakull2006} have shown that the bubbles have supersonic expansion 
speeds of 80--250 km/s, derived from the width of H\,$\alpha$, which suggests that 
(at least some parts of) the bubbles are shock-excited by winds or jets 
rather than photoionized by X-rays from the central ULX
(see also \citealt{Siwek2017}). They estimated
that about $10^{39-40}$\,erg/s is required to ionize the optical bubbles.
This is comparable to the estimates of wind kinetic power 
obtained in ULXs so far 
\citep{Pinto2016nature,Pinto2017,Walton2016a,Kosec2018a,Kosec2018b}:
\begin{flalign}\label{Eq:wind_power}
\begin{aligned}
L_w &= 0.5 \, \dot{M}_w \, v_w^2 &&\\
                 &= 2 \, \pi \, m_p \, \mu \, \Omega \, C \, \frac{v_w^3}{\xi} \, L_{\rm ion} 
                 \, \sim 10^{39-41} \, {\rm erg/s}
\end{aligned}
\end{flalign}
where $\dot{M}_w = 4 \, \pi \, R^2 \, \rho \, v_w^2 \, \Omega \, C$ is the outflow rate, $\Omega$ 
and $C$ are the solid angle and the volume filling factor (or \textit{clumpiness}), respectively,
$\rho$ is the density and $R$ is the distance from the ionizing source.
Here we have used Eq.\,(\ref{Eq:ionization_parameter}) to get rid of the $R^2 \rho$ factor
where $\rho=n_{\rm H} \, m_p \, \mu$ with $m_p$ the proton mass and $\mu = 0.6$ the 
average particle weight of a highly ionized plasma.
A solid angle $\Omega / 4\pi = 0.3$ and volume filling factor $C=0.3$ were chosen 
as fiducial average values from MHD simulations of winds driven by radiation pressure in
super-Eddington winds \citep{Takeuchi2013}.

{The filling factor of the wind might actually be much smaller, at least for the cool
and outer phases. Using Eq. (23) in \citet{Kobayashi2018} and assuming that the outflow
is comparable to the accretion rate, we obtain $C\sim5\times10^{-3}$, in which
case the wind power would be much lower, although still significantly affecting the surrounding medium.}

Following the approach of \citet{Pakull2006}, we know that the maximum mass ejected 
in the wind, $\dot{M}_w\,\tau$, throughout the ULX life time $(\tau)$ cannot be larger than 
the total mass transferred from the companion star to the Roche lobe of the compact object, 
which can be expressed as $\delta m\,M_{\odot}$, where $\delta m$ is a factor of a few
(for a standard high-mass companion star). 
From the wind kinetic luminosity $L_w$ in Eq.\,(\ref{Eq:wind_power}) we can derive
the wind velocity as 
\begin{equation}\label{Eq:wind_speed}
v_w = \sqrt{2 \, L_w \, / \, \dot{M}_w} \approx 0.19c \, \sqrt{L_{39} \, \tau_6 \, / \, \delta m}
\end{equation}
where $L_{39}$ is the wind kinetic luminosity in units of $10^{39}$ erg/s,
$\tau_6$ is the ULX life time in $10^6$ yr and {$\delta m$ is the mass in solar mass units}. 
This broadly agrees with the typical ULX 
wind velocities detected so far.

Alternatively, following the approach of 
\citet{Castor1975, Weaver1977, Soker2004, Begelman2006},
we can determine the expected radius of a bubble that has been inflated by a ULX wind. 
Assuming energy conservation, the bubble radius can be approximated as 
\begin{equation}\label{Eq:bubble_radius}
R_{\rm bubble} = 0.76 \left ( \frac{\dot{M}_w \, v_w^2}{2 \, \rho_{\rm \footnotesize{\, ISM}}} \right )^{1/5}  \tau^{3/5}
\end{equation}
where $\rho_{\rm \footnotesize{\, ISM}} \sim 10^{-24}$ g cm$^{-3}$ is the interstellar density. 
If we adopt a wind kinetic power of a few $10^{39}$ erg/s, we obtain bubble with sizes of about 
100 pc, which is consistent with the measurements quoted above.

These results have to be taken with some caution due to several uncertainties in the properties
of both the winds and the bubbles. 
For instance, we adopt a lifetime of $10^6$ yr from the 
size $\sim100$\,pc, the current expansion velocities of $\sim100$\,km/s and the
expansion low $R \sim t^{3/5}$, but it is possible
that in the past the bubbles were expanding faster.
This would correspond to shorter lifetimes.  We have adopted solid angles of $0.3$, according to MHD simulations, but \citet{Begelman2006} suggest slightly lower values ($\Omega/4\pi=0.1$). 
Finally, the total mass transferred from the companion
to the compact object might be smaller than a solar mass ($\delta m\,M_{\odot}<1\,M_{\odot}$).
However, these uncertainties would tend to cancel each other out in the equations above. 
Our predictions of radii for the wind-inflated bubbles and of wind velocities 
are likely correct within the order
of magnitude, which makes the comparisons meaningful
and therefore still likely provide useful points of comparison. 
This would provide further evidence on the detection of winds in ULXs and their role
on the surrounding environment. 

Currently, the ULX wind sample is very small due to instrumental limitations and short exposure times. 
Once a larger sample would be available it will be of high interest to compare the presence of winds in ULXs
with that of superbubbles around them -- and their characteristics -- in order to better understand their formation.

\subsection{Effects from inclination and accretion rate on the appearance of ULX winds}
\label{sec:effects_from_incl_rate}

Fig.\,\ref{Fig:plot_geometry} shows an artistic impression of a wind
launched from an accretion disc due to high radiation pressure.
{As mentioned before, this simplified geometry would apply
to the cases where the compact object is a black hole or a non-magnetic neutron star
(with the maximum magnetic field 
depending on the accretion rate).}
The wind gives the system a funnel geometry and scatters, i.e.
geometrically beams, the X-ray photons coming from the inner regions
(see, e.g., \citealt{Poutanen2007, Middleton2016}).
Therefore, we expect to see a hard X-ray spectrum followed by high 
ionization in the wind when the source is viewed at low
inclinations (i.e. on-axis) and the more the inclination increases, the softer
the X-ray spectrum and the lower the ionization of the wind should be,
for a given accretion rate (e.g., \citealt{Kawashima2012, Pinto2017}).

\begin{figure}
  \includegraphics[width=1\columnwidth, angle=0]{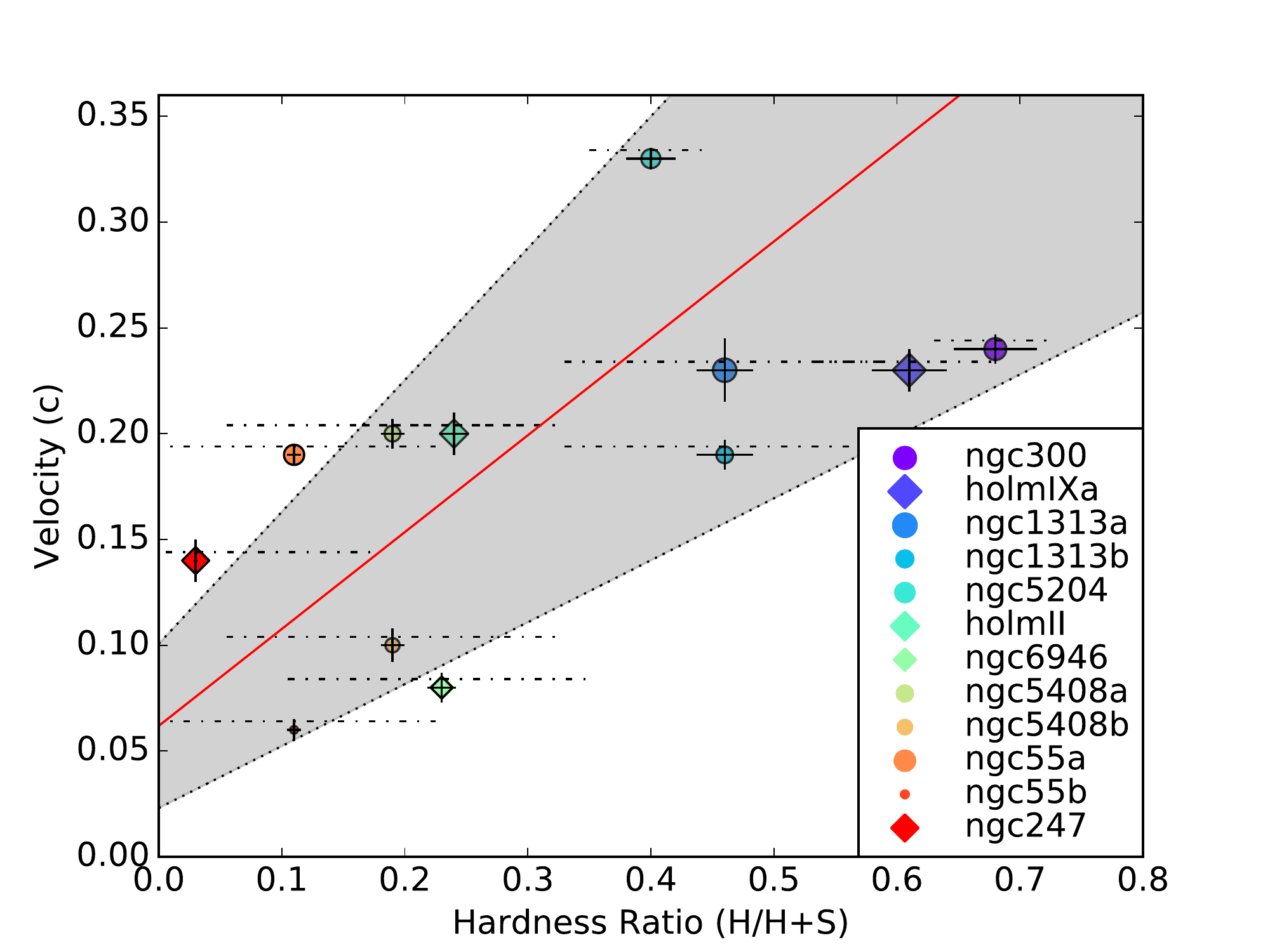}
  \includegraphics[width=1\columnwidth, angle=0]{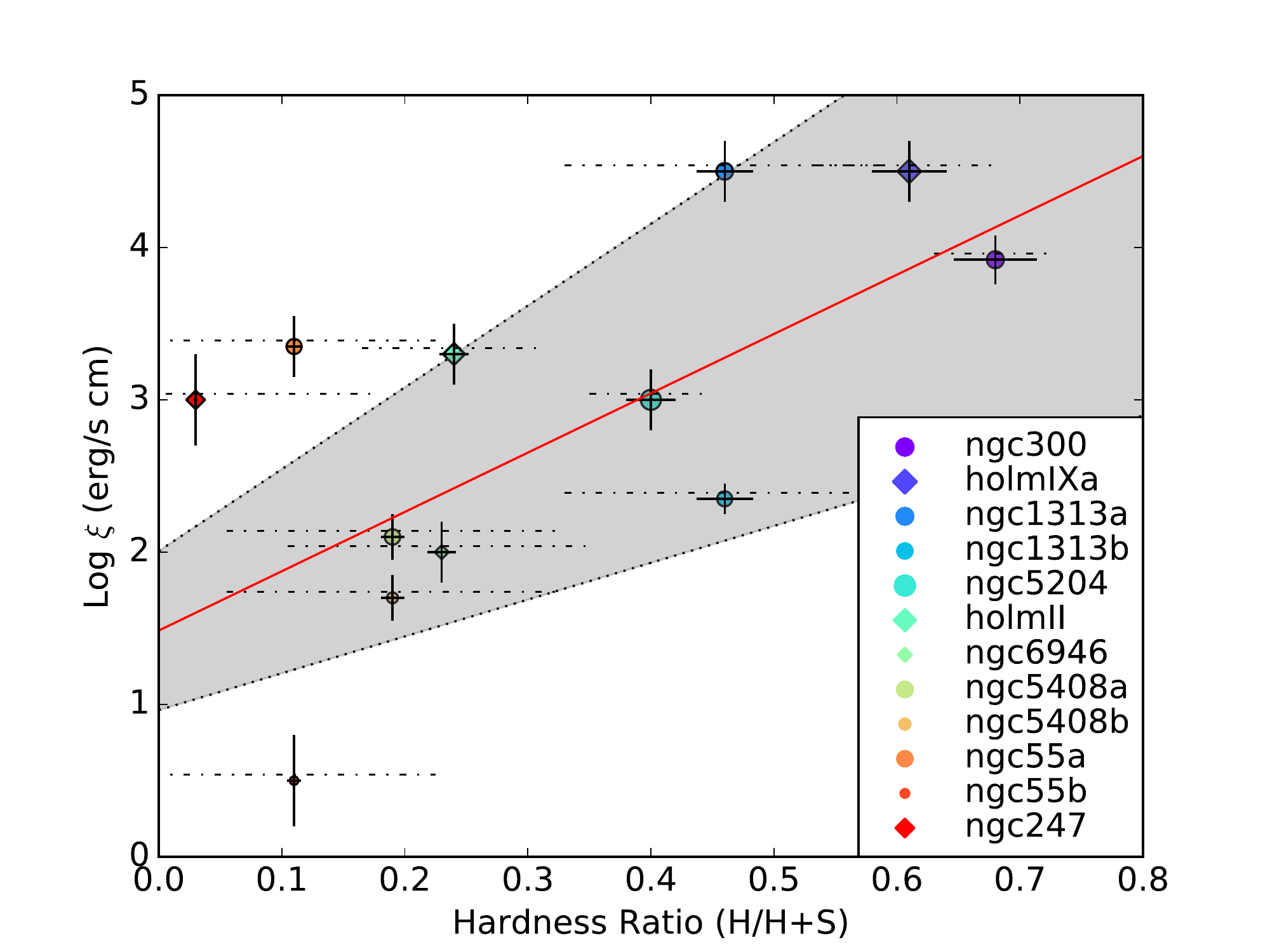}
\vspace{-0.4cm}
   \caption{Trends of wind velocity (top) and wind ionization (bottom)
                 versus spectral hardness in ULXs.
                 Point-size and colour are coded according to the ionization parameter (top)
                 and velocity of the wind (bottom). 
                 Circles and diamonds refer to significant $(>3\,\sigma)$ detections
                 and tentative $(<3\,\sigma)$ detections, respectively.
                 The shaded grey areas indicate the $1\,\sigma$ uncertainties
                 of the best-fitting straight line (see the solid red lines).
                 The larger dotted horizontal bars show the spectral hardness variability
                 of each ULX.}
   \label{Fig:plot_wind_par}
\vspace{-0.2cm}
\end{figure}

We show the trend between the velocity and the ionization parameter 
of the winds with the spectral hardness in Fig.\,\ref{Fig:plot_wind_par}. 
We notice that a third of these detections were marginal ($\lesssim3\sigma$,
see those shown as diamonds)
and that the results have to be taken with caution.
We retrieve all the necessary information on the wind properties from 
previously reported detections and double-check the results with the newer
{\scriptsize{SPEX}} version used in this paper. More detail on the wind
data is reported in Sect.\,\ref{sec:appendix_winds} and, particularly, in
Table\,\ref{table:ULX_winds}. We find a possible correlation between
the ULX spectral hardness, $HR$, the wind ionization parameter, $\xi$, and 
velocity, $v_w$. We fit a straight line to the data using 
the {\scriptsize{PYTHON}} routine 
\textit{odr}\footnote{https://docs.scipy.org/doc/scipy/reference/odr.html}
that performs orthogonal distance regressions and obtained the following 
best fits:
\begin{flalign}\label{Eq:wind_hr_trends}
\begin{aligned}
  v_w   &= (0.06 \pm 0.04) + (0.46 \pm 0.15) \, HR  &&\\
\log \xi &= (1.5 \pm 0.5) + (3.9 \pm 1.5) \, HR
\end{aligned}
\end{flalign}
where $HR$ is the hardness ratio.  
The Spearman and Pearson correlation coefficients range between 0.6 and 0.8 
(see Table\,\ref{table:correlations}), which indicates a substantial
positive correlation among the three parameters.
Although there is significant scatter, 
the apparent trend is qualitatively consistent with the scenario 
for which we see hotter and faster wind
component at lower inclination. The velocity trend may suggest that 
the optically-thin component of the wind (which is responsible for the
blueshifted absorption lines) is likely not moving along 
the equatorial plane or, more simply, that at larger inclinations we see
material expelled from an outer region where the escape velocity
is naturally lower. The ULX wind sample is currently too small to 
obtain conclusive results, but our measurements agree with the general
picture.

Some of the scatter is likely due to systematics. Here we show the 
$(HR, \xi, v_w)$ values for the observations with statistics good enough to enable
wind detections, but ULXs vary in hardness ratio.
Therefore, we also show the $HR$ fractional variability in Fig.\,\ref{Fig:plot_wind_par} 
(dotted horizontal lines) as retrieved from the literature
(\citealt{Middleton2015a, Kosec2018a, Kosec2018b, Pinto2017, Carpano2018} 
and references therein).
Although, it might not be physically correct to fit a model to these points, 
the $HR$ variability could still explain why they deviate from a tight correlation.

For the several reasons discussed above, 
the hard NGC 300 ULX-1, the intermediate NGC 1313 X-1
and the soft NGC 5408 X-1 ULXs are potentially being seen from increasing 
viewing angles. The comparison of their thermal stability curves 
in Fig.\,\ref{Fig:plot_xiP_300_5408_1313} would then confirm that 
the inner portions of the wind have to be necessarily more ionized 
and located along the $\xi-\Xi$ Compton cooling arm
or they would disappear due to strong instabilities as often observed 
in the hard states of Galactic X-ray binaries \citep{Chakravorty2016}.
%%%This seems to be consistent with the fact that harder ULX spectra 
%%%are followed by the winds detected at higher $\xi$
%%%as can be seen in Fig.\,\ref{Fig:plot_xiP_300_5408_1313}
%%%(bottom panel).
%%%This is a tentative trend since we still only have 
%%%a very small number of sources. 
%%% how about making a plot (xi,Vout,Gamma) (by Dom)? 
%%% also, could the jet die if polar winds become unstable a
%%% at high luminosities or wind densities (by ROberto)? 

Super-orbital periods have been reported in several ULXs
(see, e.g., \citealt{Kaaret2006, Motch2014, Walton2016b}). 
These may be evidence of precession which suggests that
different portions of the accretion disc and the wind would be
visible to us at different times
(see, e.g., \citealt{Luangtip2016, Middleton2018}). 
Variability in the accretion rate can cause similar behaviour
because the opening angle of the wind
is expected to be related to the accretion rate (\citealt{Poutanen2007}).
ULXs that have hard spectra at moderately-high accretion rates 
(dashed line in Fig.\,\ref{Fig:plot_SED_Balance_ngc1313_variability}, top) could potentially
undergo substantial obscuration of the inner region and appear softer
after a large increase of accretion rate (dotted line). 
In principle, this should also strengthen the soft emission due to the outer disc 
emission (dotted line). At even higher rates or inclinations
the ULX would look like a supersoft source 
(solid line).

In summary, either precession or variable accretion could expose our LOS to a 
wind portion whose ionizing field may be dominated by a region of the
accretion disc with different average temperature. 
In Sect.\,\ref{sec:ngc1313_vs_variability}, we have used the spectral
variations of NGC 1313 X-1 to understand the effects of dramatic changes in the SED.
When the source brightens from an intermediate-hard (HUL) state to  
the bright-broadened-disc state (BBD) and reaches its X-ray luminosity peak,
the $S$ curve changes, extending the region of instability and, therefore, 
affects the wind ``warm'' component with low ionization parameter  
($\log \xi \sim1.8-2.3$, dotted line in 
Fig.\,\ref{Fig:plot_SED_Balance_ngc1313_variability}, bottom). 
This means that this wind component is either not in equilibrium or simply out
of our LOS, in each case basically undetectable. This can explain the
weakening of the 1 keV residuals previously seen in \citet{Middleton2015b}
and seems to be confirmed by new, deeper, XMM-\textit{Newton} observations of 
NGC 1313 X-1 \citep{Pinto2019b}. 

\subsection{Caveats}
\label{sec:caveats}

It is worth noticing that the black holes powering AGN - which accrete at much lower 
Eddington ratio than those in ULXs - exhibit ultrafast outflows within a broad range 
of inclination angles and their wind parameters
do not show strong correlations with disc characteristics such as the
inclination (see, e.g., \citealt{Tombesi2014}). This would suggest that
AGN winds might differ from those seen in highly super-Eddington sources like ULXs,
possibly due to a different launching mechanism. In fact, low-Eddington AGN winds
are likely driven by magnetic pressure (see, e.g., \citealt{Fukumura2017}
and references therein).

We also address briefly the possibility that some 
portions of the wind can be launched by thermal heating
close to the Compton temperature (see, e.g., \citealt{Begelman1983, Done2018}).
In our case, the Compton radius will be $R_{IC}=GM\mu m_{\rm p}/(kT_{IC})\sim3\times10^5R_S$ 
for a $10\,M_{\odot}$ black hole and $T_{IC}\sim10^7$\,K (i.e. the temperature where 
the $T-\xi$ curve flattens for the ULX in Fig.\,\ref{Fig:plot_SED_Balance}, middle panel).
This is much larger than the spherization radius
$R_{\rm \, sph}=5/3 \dot{M} k / (8 \pi c) \sim 50R_S$ and the photospheric radius
$R_{\rm \, phot}=\dot{M}_w k / (4 \pi v_w \cos\theta) \sim 500 R_S$, 
the latter indicating where the wind becomes optically thin.
Here we adopt $\dot{M}=10\dot{M}_{\rm Edd}$, $\dot{M}_w=\dot{M}$, $v_w=0.2c$, 
$\theta=45^{\circ}$ and $\dot{m}=10$ 
(see, e.g., \citealt{Poutanen2007, Fiacconi2017}).
The hard X-rays from the inner region will likely be screened by the thick disc atmosphere
or the wind itself, and it will be difficult for them to reach the putative
Compton radius (just like qualitatively shown in Fig.\,\ref{Fig:plot_geometry}).
Therefore, we do not expect to see a classical Compton wind with slow velocities 
($\sim500$\,km/s) from the outer disc.

{A word of caution is required regarding the velocity of the wind for magnetic accretors. 
The current estimates suggest some winds as fast as $0.2c$ (see e.g. Fig.\,\ref{Fig:plot_wind_par}). 
If this corresponds to the escape velocity, for a neutron star with $B\sim10^{12}$\,G, 
it would mean that the wind has to be launched from inside the magnetosphere 
($R_{M} \sim 100 R_S$), with possible additional contributions from the magnetic fields themselves. 
This would not be the case for much lower magnetic fields, e.g. $B\lesssim10^{10}$\,G, 
or if the detected high velocities are due to later vertical 
acceleration (e.g. \citealt{Takeuchi2013}). Further observations and comparisons
of winds in pulsating vs. non-pulsating ULXs might help to understand the phenomenology.}

{Another caveat for the applicability of our model and calculations of thermal stability 
concerns the recombination timescales of the plasma. If such timescales are much longer 
than the timescale of the expansion necessary to drastically change the ionizing field then
the gas might be out of equilibrium. The recombination timescales depend on the specific ion,
they tend to be longer for higher ionization states. Through the \textit{pion} model in {\scriptsize{SPEX}}
it is possible to calculate the product between the electron density $n_{\rm e}$ and the 
recombination time $t_{\rm rec}$ for a given SED. Using the time-averaged SED of NGC 1313 X-1
(see Fig.\,\ref{Fig:plot_SED_Balance}, top panel) and adopting $\log \xi = 2.3$ from the best fits 
(see Table\,\ref{table:ULX_winds} and \citealt{Pinto2016nature}), we estimate
$n_{\rm e} \cdot t_{\rm rec} = 
n_i / (n_{i+1} \alpha_{i+1} - n_{i} \alpha_{i}) \sim 10^{9-10} {\rm cm}^{-3}  {\rm s}$,
where $n_{i}$ is the density of the ion in the state $i$ and $\alpha_{i+1}$ is the recombination rate 
coefficient (in ${\rm cm}^{3}  {\rm s}^{-1}$) from state $i+1$ to $i$.
The range indicates the values calculated for the most relevant transitions of the following ions:
{O\,{\sc vii-viii}} and {Ne\,{\sc ix-x}}. \citet{Pinto2019b} placed a constraint on the density 
of the gas responsible for the emission through the {O\,{\sc vii}} triplet: 
$n_{\rm H} \sim 10^{10-12} {\rm cm}^{-3}$ (with $10^{11} {\rm cm}^{-3}$ as a median value) 
similar to disc coronae in Galactic X-ray binaries. 
The outflowing gas coming from the inner regions might be at even higher 
densities. This would suggest $t_{\rm rec}\sim 0.001-1 {\rm s}$ or shorter 
(with a median value around $0.05 {\rm s}$). On average we would expect the gas to have 
travelled around 2 $R_{\rm \, sph}$ or 0.2 $R_{\rm \, phot}$ during the recombination time scale
(for $v=0.2c$, $\dot{M}=10\dot{M}_{\rm Edd}$ and $M_{\rm BH} = 10\,M_{\odot}$). 
In the most pessimistic case it would have travelled a few $R_{\rm \, phot}$,
where the photospheric radius is the region with optical depth close to unity.
We expect a similar condition for a neutron star with 
an accretion rate of $100\dot{M}_{\rm Edd}$.
This means that there can be enough time for the outflowing gas to recombine before 
dramatical changes would happen in the ionizing field.} %%% add mass and mdot of BH adopted ...

%%%A time scale of several seconds, in particular, corresponds to a radius a couple of orders of magnitude larger than the spherization radius where the wind supposedly originates from. If the gas actually observed resides at R>~ 10^{10}cm, where is the contribution of the gas in the distance range 10^8..10^{10}cm that apparently should be a couple of orders of magnitude larger, as ? ? 1/R? And the ionization and temperature of this gas
%%%would be non-equilibrium if the recombination and cooling time scales are several
%%%seconds. Anyway, I do not insist on a complete analysis, but rather adding an
%%%approximate estimate of the range of radii and the possible physical conditions where the model may be applied.

{In this paper we have shown a first step towards the characterization and the 
understanding of winds in super-Eddington accretion discs of stellar-mass black holes.
In the future it will be important to further investigate systematic effects due to relativistic
corrections due to the velocity of the wind and to time scales of its expansion as
compared to the recombination and cooling time scales.}

%%%$R_{\rm phot}=\dot{M}_w k / (4 \pi v_w \cos\theta) \sim 300R_S$ or 
%%%$R_{\rm phot}\sim 3 \, \epsilon_w \, \dot{m}^{3/2} $
%%%and $\epsilon_w = L_w / L_{tot} > 0.5$
%%%or a minimum 50\% energy loss through the wind

\subsection{Future missions: \textit{XRISM} and \textit{ATHENA}}
\label{sec:discussion_missions}
 
XMM-\textit{Newton} and \textit{Chandra} have provided strong insights in the knowledge 
of ULXs and accretion physics in general since their launches 20 years ago. 
However, newly planned missions will significantly
improve our ability to detect winds in these objects.
\textit{XRISM}\footnote{X-Ray Imaging and Spectroscopy Mission,
a.k.a \textit{XRISM}, to be launched in 2022, see https://heasarc.gsfc.nasa.gov/docs/xrism.}, 
for instance, will enable for the first time observations of ULXs with a
good collecting area and an outstanding spectral resolution of about 5\,eV 
in the hard X-ray band ($\sim2-10$\,keV), a combination missing in current telescopes
 \citep{Guainazzi2018}.

\begin{figure}
  \includegraphics[width=1\columnwidth, angle=0]{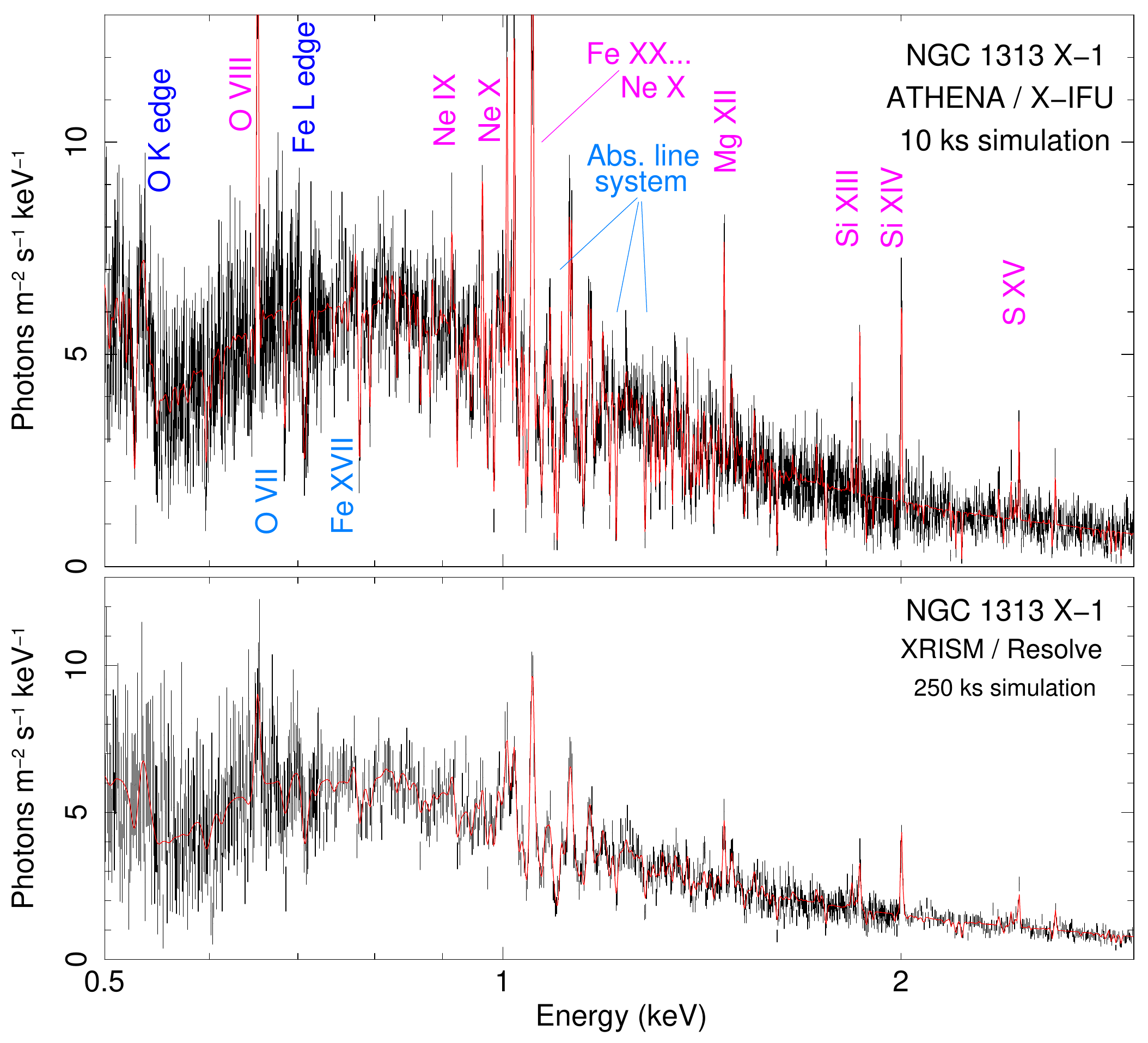}
   \caption{Simulations of the ULX NGC 1313 X-1
                 adopting the XMM-\textit{Newton}/RGS best fit model
                 from \citet{Pinto2016nature}.
                 Cosmic X-ray background and particle background are accounted for.
                 The strongest transitions are labelled in the observed frame.
                 We use the response matrix of {\textit{Hitomi}}/SXS as a proxy for {\textit{XRISM}}.}
   \label{Fig:plot_ATHENA_XRISM}
%%%  \vspace{-0.1cm}
\end{figure}

Here we perform a simulation of NGC 1313 X-1 as it may be observed by 
{\textit{XRISM}} using its predecessor {\textit{Hitomi}}/SXS 5\,eV response matrix 
(see Fig.\,\ref{Fig:plot_ATHENA_XRISM}, bottom panel).
We adopt the best fit model from \citet{Pinto2016nature}
consisting of a spectral continuum (blackbody $T=0.3$ keV and powerlaw 
$\Gamma=1.9$ components), an isothermal rest-frame collisionally-ionized
plasma with a temperature of 0.8 keV, an outflowing photoionized plasma
(column density $N_{\rm H}=5\times10^{21}{\rm cm}^{-2}$, 
$\log \xi=2.3$, $v_{\rm LOS} = 0.2c$, $v_{\rm turb} = 500$ km\,s$^{-1}$),
and Galactic absorption through the $hot$ model in {\scriptsize{SPEX}}
with $N_{\rm H}=2\times10^{21}{\rm cm}^{-2}$ (see, e.g., \citealt{Pinto2013}).
The simulation performed with the {\textit{XRISM}}/Resolve microcalorimeter
provides a clear view of the Fe L region around 1 keV and boosts the detection
of lines above 2 keV with the resolution necessary to break degeneracy
such as between the ionization parameter and the outflow velocity.
However, we will still need observations of 100\,ks or above in order to get 
$5\,\sigma$ significance detections for the strongest lines.
Below $\sim0.7$\,keV the XMM-\textit{Newton}/RGS grating spectrometer
will still be the most sensitive instrument.

X-ray astronomy will radically change in the 2030s after the launch of ATHENA 
(\citealt{Nandra2013, Guainazzi2018}). The mission will have both a wide 
field imager (WFI) with collecting area an order of magnitude above any current 
X-ray mission over the $0.3-10$ keV energy band and a field of view of 40'$\times$40',
along with the X-ray Integral Field Unit (X-IFU). The X-IFU will be 
the first X-ray microcalorimeter to provide both high spatial (5'') and spectral (2.5 eV)
resolution with an effective area an order of magnitude higher
than any current high-spectral resolution X-ray detector. This corresponds
to an improvement of almost 2 orders of magnitude in the way we detect and resolve
narrow features in the canonical 0.3--10 keV X-ray energy band.

We perform another simulation of NGC 1313 X-1 using the same
baseline model but only assuming an exposure time of 10\,ks and the response
matrix of ATHENA/X-IFU (dated late 2018). The expected results are extremely promising
(Fig.\,\ref{Fig:plot_ATHENA_XRISM}, top).
Even a short exposure with X-IFU will provide a high signal-to-noise 
spectrum with a forest
of emission and absorption lines all over the soft X-ray band, 
a spectrum quality comparable to the XMM-\textit{Newton}/RGS spectra 
of very bright Galactic novae (see, e.g., \citealt{Ness2011}), 
with the remarkable difference that the ULX is located 4 Mpc away, 
while novae are within a few kpc from us.

The X-IFU spectrum will detect dozens of emission and absorption lines
well above $5\,\sigma$ each and provide strong constraints on emission lines from triplets,
such as Ne\,{\scriptsize{IX}}, and other lines sensitive to plasma density.
More importantly, we will be able to detect winds in ULXs at much larger distance,
therefore boosting the explored volume (dozens of ULXs with winds detections) 
and the wind parameter space including its energetics and variability.
This will be crucial to fully address the relationships between the spectral
hardness of the source and the wind properties already pointed
in Eq.\,(\ref{Eq:wind_hr_trends}) and Fig.\,\ref{Fig:plot_wind_par}.
It is also worth mentioning that the X-IFU will enable, for the first time, short-term
variability studies of ULX winds and their connections with X-ray lags.

\section{Conclusions}
\label{sec:conclusion}

In this work we have provided the first insights on the thermal stability of winds in 
ultraluminous X-ray sources (ULXs), currently the best candidates of astronomical
persistent sources to accrete beyond the Eddington limit. 
{We focus on the case where the compact object is either a black hole or 
a non-magnetic neutron star for which a thick disc would still form and launch
powerful winds through radiation pressure}.
Using three archetypal ULXs
with wind detections, we have probed the thermal stability for three strategic spectral states 
that broadly describe the ULX phenomenology: soft, hard and broadened-disc
state. We have found that winds in ULXs are likely in stable thermal equilibrium
with the harder states progressively extending the unstable branch.
The stability curves resemble those of Narrow Line Seyfert 1 and supersoft sources
as expected at high-Eddington rates. Our results predict the existence of the
evacuated funnel in ULXs as the wind is unstable and/or highly-ionized 
in innermost regions, likely providing a better view of the hard X-ray region.

We have also investigated the consequence of X-ray spectral variability and, in general,
dramatic variations in the ionizing field by studying some ad-hoc spectral states of ULX 
NGC 1313 X-1.
The results have confirmed that over the periods of time where strong variability
occurs (typically weeks-to-months in ULXs) the thermal stability also significantly varies, 
which can explain the disappearance of strong residuals 
(and therefore winds) during bright states. 
Precession and accretion rate changes could both contribute to this.
We also find a possible correlation between the spectral hardness of the ULX,
the wind velocity and the ionization parameter in support of the overall scenario
and the geometry of super-Eddington accretion discs.

%%%If a significant fraction ($\gtrsim10$\,\%) of the optical-UV light is self-screened 
%%%by the disc thick atmosphere and the wind,
%%%and it cannot access the wind inner regions, then the unstable branch extends 
%%%towards lower values of the ionization parameter, $\log \xi$. 
%%%This forces the wind out of intermediate values, e.g. $\log \xi\sim2.0-2.5$,
%%%with a consequent suppression of the features around 1 keV as observed 
%%%in many ULXs and confirmed by recent results.

\appendix

\section{Technical detail}
\label{sec:appendix}

Here we show some technical information that we avoid in the main body of the paper
to facilitate the reading. 

\subsection{Observation log}
\label{sec:appendix_obs}

In Table\,\ref{table:obs_log}, we report the XMM-\textit{Newton} 
archival observations used to build the SEDs. In particular, the three longest observations taken during the intermediate-hard state where NGC 1313 X-1 spends most time and
two specific observations yielding the softest and brightest state recorded
for this source.
The latter two have been used in Sect.\,\ref{sec:ngc1313_vs_variability} to compute 
the ionization balance of NGC 1313 for different spectral shapes
(see Fig.\,\ref{Fig:plot_SED_Balance_ngc1313_variability}).
It is remarkable how NGC 1313 X-1 reaches extremely faint, soft states 
(e.g. for observation 0205230401) where its hardness ratio agrees (within error bars)
with the brightest state of the supersoft ULX NGC 247 (see Table\,\ref{table:ULX}).
This is of course expected from the unification scenario 
owing to a combination of lower accretion rates and system precession
(see \citealt{Middleton2015a} for more detail on spectral variability of ULXs).

{The other ULX observations and RGS spectra used to produce the results shown
in Table \ref{table:ULX} and \ref{table:ULX_winds} and in Fig. \ref{Fig:plot_wind_par} 
are the following: 
0200980101 (Holmberg IX),
0200470101 (Holmberg II), 0691570101 (NGC 6946 X-1), 
0142770101-301, 0405690101-201-501, 0693850701-1401 
and 0741960101 (NGC 5204, e.g. \citealt{Kosec2018a}),
0655050101 (NGC 55) and 0728190101 (NGC 247, e.g. \citealt{Pinto2017}).}

\subsection{Heating and cooling rates} %%% should be check with new pion / correct SED
\label{sec:appendix_rates}

\begin{figure*}
  \includegraphics[width=1\columnwidth, angle=0]{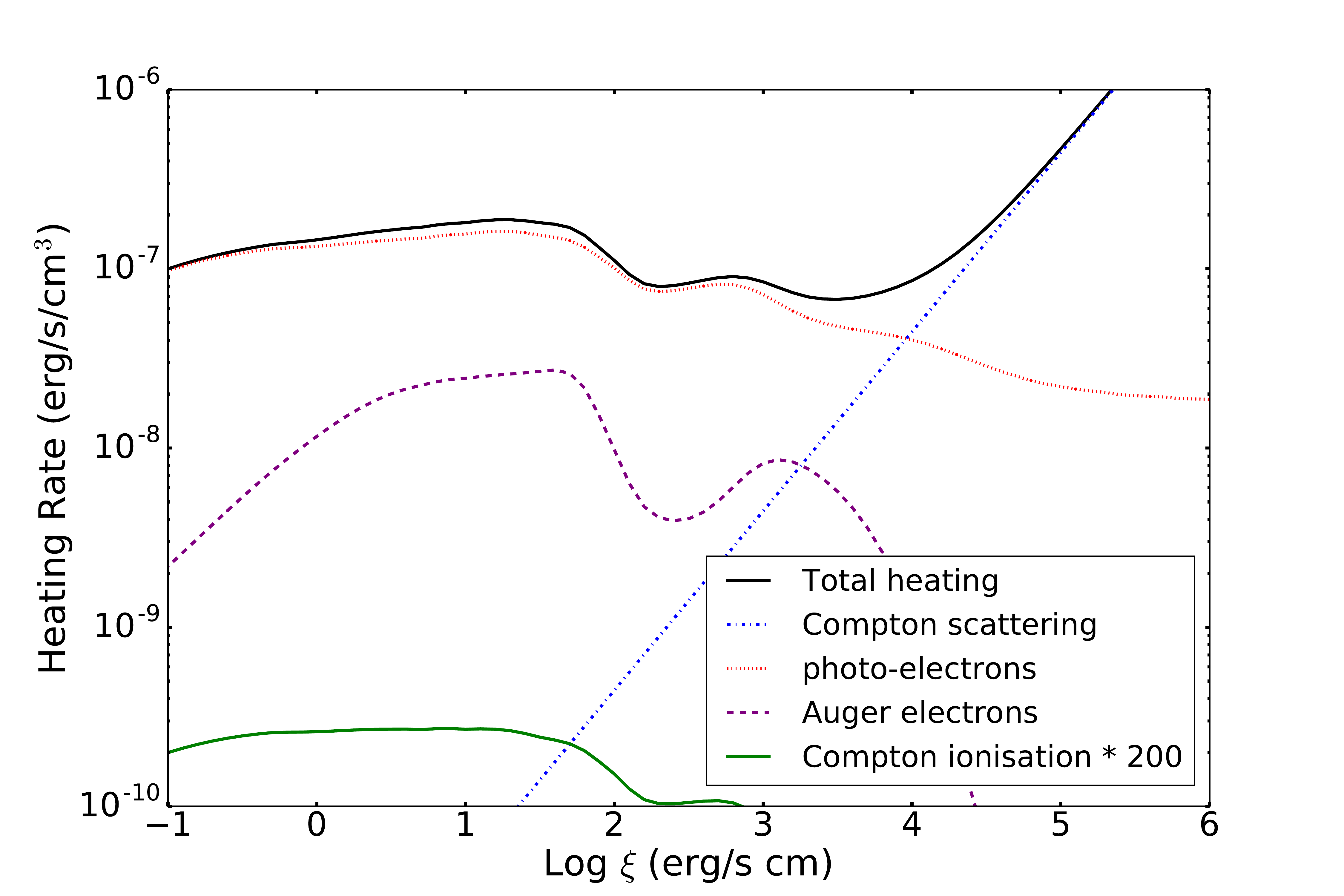}
%  \hspace{0.3cm}
  \includegraphics[width=1\columnwidth, angle=0]{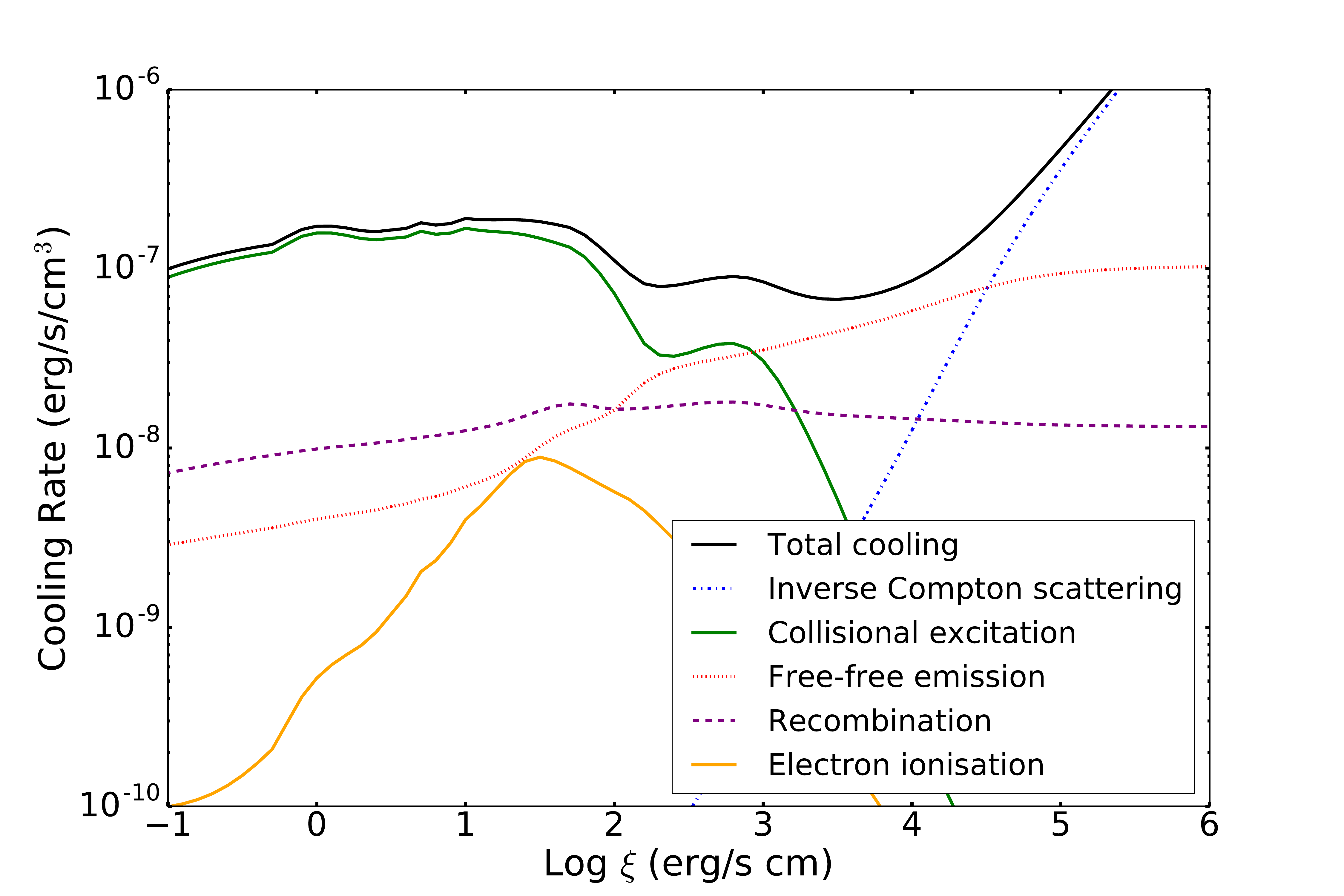}
   \caption{Heating rates (left panel) and cooling rates
                (right panel) calculated for NGC 1313 X-1 with the
                 {\scriptsize{SPEX}} \textit{pion} code.}
   \label{Fig:plot_rates}
\end{figure*}

In Fig.\,\ref{Fig:plot_rates} we show the heating and cooling rates computed 
for the broadband SED of NGC 1313 X-1 using the intermediate-hard time-average
spectrum (see Sect.\,\ref{sec:SED}, Fig.\,\ref{Fig:plot_SED_Balance} and 
Table\,\ref{table:obs_log}).
These rates are computed using the {\scriptsize{SPEX}} \textit{pion} code.
These plots help to understand the behaviour of different physical processes at different 
ionization parameters, i.e. in different wind conditions. 
For our calculations, we set the hydrogen density to $n_{\rm H} = 10^{8}$ cm$^{-3}$ 
and the total column density to $N_{\rm H} = 10^{22}$ cm$^{-2}$
similarly to \citet{Mehdipour2016}. We notice that in the regime of non-optically-thick gas
the main effect of the density is a systematic scaling of all heating and cooling rates.

At first, we notice that the total heating and cooling rates are broadly consistent,
which means that a converging, equilibrium solution is found at nearly any 
value of the ionization parameter, $\xi$. At low $\xi$ the dominant processes
are heating by photo-electrons and cooling by collisional excitation.
At high $\xi$ the largest contribution to heating is given by Compton scattering
and to cooling by inverse Compton scattering.
There is therefore a radical change in the dominant physical process with the
ionization parameter and, therefore, the wind temperature.
This is consistent with what \citet{Mehdipour2016} found comparing 
blackbody-like SEDs to harder spectra of AGN where other processes,
such as auger electrons and recombination, provide more significant contribution 
to heating and cooling, respectively.
They also compared the results obtained by 
{\scriptsize{SPEX}} \textit{pion} with those using different spectral codes 
such as {\scriptsize{XSTAR}} and {\scriptsize{CLOUDY}}, finding some interesting
differences of 10--30\% in the corresponding estimates of $\xi$ and the 
concentration of several ionic species. However, they argue that
these differences are unlikely to impact the scientific interpretation 
of current (AGN) outflow observations.

\subsection{Wind properties in ULXs} %%% should be check with new pion / correct SED
\label{sec:appendix_winds}

In Table\,\ref{table:ULX_winds} we quote the hardness ratios of each source
calculated as $HR=L_{2-10 \rm keV} / L_{0.3-10 \rm keV}$ using 
a $bb+mbb+pow$ spectral continuum 
(soft blackbody + broad modified
blackbody + powerlaw, see e.g. \citealt{Kosec2018b})
for the XMM/EPIC-pn spectra from observations with evidence of winds 
(see Fig.\,\ref{Fig:plot_ULX_sequence} and Sect.\ref{sec:ULX_intro}). 
We also report the velocities and the ionization parameters of the 
wind along with the references used. These data are used to 
produce Fig.\,\ref{Fig:plot_wind_par}.

For NGC 5204 X-1 we fit the data with the new $pion$ model in emission
adopting $\Omega=2\pi$ in order to obtain a measurement of the ionization
parameter (see, e.g., \citealt{Psaradaki2018}). 
This is done because \citet{Kosec2018a} show spectral fits using 
only line-emission from gas in collisional equilibrium, which reproduces the lines
better than photoionization equilibrium (but deeper observations are necessary to 
distinguish between a collisionally-ionized and a photo-ionized nature of the outflow). 

For NGC 1313 X-1 we use the 
fastest and the slowest solutions found by \citet{Pinto2016nature} 
and \citet{Walton2016a} because they yield the best description of the soft and hard
X-ray components of the wind and also provide an idea of the systematic uncertainties. 

We double-check the values for Holmberg II X-1, Holmberg IX X-1 and NGC 6946 X-1
by fitting the RGS+EPIC/pn spectra used in \citet{Kosec2018a} and \citet{Pinto2016nature}. 
Here, we report the results obtained for the solution that 
correspond to the highest significance. We still quote the wind detection in these
sources as tentative because the Monte Carlo simulations yield a significance below
$3\,\sigma$, likely owing to the limited statistics. 
The previous results have all been double-checked by testing the new version
of SPEX used in this paper (3.05.00).

In Table\,\ref{table:correlations} we quote the Spearman and Pearson correlation 
coefficients for the relationships between
the ULX spectral hardness, $HR$, the wind ionization parameter, $\xi$, and 
velocity, $v_w$. 
They are calculated using the {\textit{scipy.stats.spearmanr}} and 
{\textit{scipy.stats.pearsonr}} {\scriptsize{PYTHON}} routines for
the dataset shown in Table\,\ref{table:ULX_winds}.
Despite the large scatter, the results indicate a substantial
positive correlation among the three parameters, which agrees with the 
predictions for super-Eddington disc geometry and winds.

\begin{table}
\caption{XMM-\textit{Newton} observations used to build ULX SEDs.}  
% \vspace{-0.2cm}
\label{table:obs_log}      % is used to refer this table in the text
\renewcommand{\arraystretch}{1.2}
 \small\addtolength{\tabcolsep}{-1pt}
 
\scalebox{1}{%
\begin{tabular}{c c c c c c c c c c c c c c c c c c}     
\hline  
Source           & State           & $HR$ & OBS\_ID & t$_{\rm exp}^{\rm tot}$\,(ks)  \\
\hline                                                                                                           
NGC\,300       &  Hard         & \multirow{ 2}{*}{0.7} &  0791010101  &  \multirow{ 2}{*}{201.0}    \\
NGC\,300       &  Hard         &                                 &  0791010301  &      \\
NGC\,1313     &  Int-Hard    & \multirow{ 3}{*}{0.5} &  0405090101  &  \multirow{ 3}{*}{345.6}    \\
NGC\,1313     &  Int-Hard    &                                 &  0693850501  &      \\
NGC\,1313     &  Int-Hard    &                                 &  0693851201  &      \\
NGC\,1313     &  Bright-BB &   0.35                        &  0301860101  &  21.3    \\
NGC\,1313     &  Low-Soft  &   0.04                        &  0205230401  &  14.3    \\
NGC\,5408     &  Soft          & \multirow{ 6}{*}{0.2}  &  0302900101  &  \multirow{ 6}{*}{644.9}   \\
NGC\,5408     &  Soft          &                                  &  0500750101  &      \\
NGC\,5408     &  Soft          &                                  &  0653380201  &      \\
NGC\,5408     &  Soft          &                                  &  0653380301  &      \\
NGC\,5408     &  Soft          &                                  &  0653380401  &      \\
NGC\,5408     &  Soft          &                                  &  0653380501  &      \\
\hline                
\end{tabular}}

Notes: Exposure times refer to the RGS\,1,2 average. 
Unabsorbed hardness ratios $(HR)$ are computed as 
$L\,_{2-10 \rm keV}$ / $L\,_{0.3-10 \rm keV}$.
The broadband SEDs are shown in  
Fig.\,\ref{Fig:plot_xiP_300_5408_1313} 
and \ref{Fig:plot_SED_Balance_ngc1313_variability}.
\end{table}

\begin{table}
\renewcommand{\tabcolsep}{1.3mm}
\renewcommand{\arraystretch}{1.2}
%\footnotesize
\caption{\label{table:ULX_winds} Wind properties and spectral hardness in ULXs} 
\begin{tabular}{c|ccc|cccc|cc}
\hline
Source                     &  $HR$     &   $v_w$   & $\log \xi$ & REF   \\ 
\hline                                              
{NGC 300 ULX-1}        &  $ 0.68 \pm 0.03 $  &  $0.24  \pm 0.01$  & $  3.9   \pm 0.2 $  & [5]    \\  
{Holmberg IX X-1}   &  $ 0.61 \pm 0.03 $  &  $0.23  \pm 0.01$  & $  4.5   \pm 0.2 $  & [4] \\   
{NGC 1313 X-1} (a) &  $ 0.46 \pm 0.02 $  &  $0.23  \pm 0.02$  & $  4.5  \pm 0.2 $  &  [1,3] \\   
{NGC 1313 X-1} (b) &  $ 0.46 \pm 0.02 $  &  $0.19  \pm 0.01$  & $  2.3  \pm 0.1 $  &  [1,3] \\   
{NGC 5204 X-1}       & $ 0.40 \pm 0.02 $  &  $0.34  \pm 0.01$  & $  3.0  \pm 0.2 $   & [4]    \\   
{Holmberg II X-1}     &  $ 0.24 \pm 0.01 $  &  $0.20  \pm 0.01$  & $  3.3 \pm 0.2  $  & [4]  \\   
{NGC 6946 X-1}       & $ 0.23  \pm 0.01 $  &  $0.08  \pm  0.01$  & $ 2.0  \pm 0.2  $  & [1,4]  \\   
{NGC 5408 X-1} (a)  & $ 0.19  \pm 0.01 $  &  $0.20  \pm  0.01$  & $ 2.1  \pm 0.2  $  & [1]  \\   
{NGC 5408 X-1} (b)  & $ 0.19  \pm 0.01 $  &  $0.10  \pm  0.01$  & $ 1.7  \pm 0.2  $  & [1]  \\   
{NGC 55 X-1}     (a)  & $ 0.11  \pm 0.01 $  &  $0.19  \pm  0.01$  & $ 3.3  \pm 0.2  $  & [2]  \\   
{NGC 55 X-1}     (b)  & $ 0.11  \pm 0.01 $  &  $0.06  \pm  0.01$  & $ 0.5  \pm 0.3  $  & [2]  \\   
{NGC 247 X-1}         & $ 0.03  \pm 0.01 $  &  $0.14  \pm  0.01$  & $ 3.0  \pm 0.3  $  & [2]  \\   
\hline
\end{tabular}

% \vspace{0.2cm}
{  Hardness ratios ($HR=L_{2-10 \rm keV} / L_{0.3-10 \rm keV}$)
   are estimated adopting a $bb+pow+mbb$ spectral continuum
   model for the XMM/EPIC-pn spectra from the observations where winds were found
   (see Fig.\,\ref{Fig:plot_ULX_sequence} and Sect.\ref{sec:ULX_intro}). 
   (a,b) refer to different wind components found in the same source.
   The wind line-of-sight velocities are in units of speed of light, $c$,
   and the ionization parameters $\log \xi$ 
   are in units of erg/s cm. The references are
   [1] \citet{Pinto2016nature}, [2] \citet{Pinto2017}, [3] \citet{Walton2016a}, 
   [4] \citet{Kosec2018a} and [5] \citet{Kosec2018b}. }
%\vspace{-0.5cm}
\end{table}

\begin{table}
\renewcommand{\tabcolsep}{2.mm}
\renewcommand{\arraystretch}{1.2}
%\footnotesize
\caption{\label{table:correlations} Wind properties and spectral hardness correlations} 
\begin{center}
\begin{tabular}{c|ccc|cccc|cc}
\hline
Parameters           & Spearman & Pearson  \\ 
\hline                                              
$HR-v_w$             &  $ 0.71 $  &   $ 0.61 $   \\  
$HR-\log \xi$    &  $ 0.58 $  &   $ 0.62 $   \\   
$v_w-\log \xi$   &  $ 0.76 $  &   $ 0.70 $    \\   
\hline
\end{tabular}
\end{center}
% \vspace{0.2cm}
{  Spearman and Pearson correlation coefficients calculated for
   the trends between Hardness ratios $HR$, wind components velocity $v_w$
   and ionization parameter $\log \xi$. 
   See also Table\,\ref{table:ULX_winds} and Fig.\,\ref{Fig:plot_wind_par}. }
%\vspace{-0.5cm}
\end{table}

%%%\begin{tabular}{c|ccc|cccc|cc}
%%%\hline
%%%$HR-v_w$   & $HR-\log \xi$ &  $v_w-\log \xi$  \\    
%%%$0.61-0.71$ & $0.58-0.62$       &  $0.70-0.76$  \\   
%%%\hline
%%%\end{tabular}

\section*{Acknowledgments}

This work is based on observations obtained with XMM-\textit{Newton}, an
ESA science mission funded by ESA Member States and USA (NASA).
CP is supported by European Space Agency (ESA) Research Fellowships.
MM is supported by the Netherlands Organisation for Scientific Research (NWO) through the Innovational Research Incentives Scheme Vidi grant 639.042.525.
ACF and PK acknowledge support from ERC Advanced Grant Feedback 340442,
the European Union's Horizon 2020 Programme 
under the AHEAD AO5 project (grant agreement n. 654215)
and the Faculty of the European Space Astronomy Centre (ESAC) 
under the project number 459.
DJW and MJM thank STFC for support in the form of an Ernest Rutherford Fellowship.
TPR acknowledges funding from the STFC via consolidated grant ST/P000541/1.
We thank Jelle de Plaa for support in optimising {\scriptsize{SPEX}},
Douglas Buisson and William Alston for support with the XMM-\textit{Newton}/OM. 
We also thank Manfred Pakull for useful discussions
on the effects of ULX winds onto the surrounding bubbles
and Ioanna Psaradaki for useful support with the {\scriptsize{PYTHON}}
and the {\scriptsize{SPEX}} \textit{pion} codes.
We finally acknowledge the anonymous referee for useful comments
that significantly improved the paper.

\bibliographystyle{mn2e}
\bibliography{bibliografia} %----> bibliografia.bib

%%%\bsp

\label{lastpage}

\end{document}